\documentclass[aps,prd,twocolumn]{revtex4}
\usepackage{amsmath,amssymb,amsfonts}
\usepackage{graphicx}
\usepackage{enumerate} 
\usepackage{colordvi} 
\usepackage{bm}

\newcommand{\be}{\begin{eqnarray}}
\newcommand{\ee}{\end{eqnarray}}
\newcommand{\simgt}{\lower.5ex\hbox{$\; \buildrel > \over \sim \;$}}
\newcommand{\simlt}{\lower.5ex\hbox{$\; \buildrel < \over \sim \;$}}

\newcommand{\bfk}{{\bf k}}
\newcommand{\bfp}{{\bf p}}

\newcommand{\deltah}{\delta_{\rm h}}
\newcommand{\deltam}{\delta_{\rm m}}

\newcommand{\Pm}{P_{\rm m}}

\newcommand{\Ph}{P_{\rm h}}
\newcommand{\Phm}{P_{\rm hm}}
\newcommand{\Pnw}{P_{\rm nw}}

\newcommand{\Omegam}{\Omega_{\rm m}}
\newcommand{\Omegab}{\Omega_{\rm b}}

\newcommand{\rmd}{{\rm d}}
\begin{document}
\title{Baryon Acoustic Oscillations in 2D II: Redshift-space halo clustering 
  in N-body simulations}
\vfill
\author{Takahiro Nishimichi$^{1}$, Atsushi Taruya$^{1,2}$}
\bigskip
\address{$^1$Institute for the Physics and Mathematics of the Universe, 
University of Tokyo, Kashiwa, Chiba 277-8568, Japan}
\address{$^2$Research Center for the Early Universe, School of Science, 
The University of Tokyo, Bunkyo-ku, Tokyo 113-0033, Japan}
\bigskip
\date{\today}
%

\begin{abstract}
We measure the halo power spectrum in redshift space from cosmological 
N-body simulations, and test the analytical models of redshift distortions 
particularly focusing on the scales of baryon acoustic oscillations (BAOs). 
Remarkably, the measured halo power spectrum in redshift space exhibits a 
large-scale enhancement in amplitude relative to the real-space clustering, 
and the effect becomes significant for the massive or highly biased 
halo samples. 
These findings cannot be simply explained by the so-called streaming model 
frequently used in the literature. By contrast, 
a physically-motivated perturbation theory model developed in the 
previous paper reproduces the halo power 
spectrum very well, and the model combining a simple linear scale-dependent 
bias can 
accurately characterize the clustering anisotropies of halos in two 
dimensions, i.e., line-of-sight and its perpendicular directions. 
The results highlight the significance of non-linear coupling 
between density and velocity fields associated with two competing effects 
of redshift distortions, i.e., Kaiser and Finger-of-God effects, and 
a proper account of this effect would be important in accurately 
characterizing the BAOs in two dimensions.
\end{abstract}

\maketitle

\maketitle
\flushbottom
\section{Introduction}
Redshift distortions caused by the systematic effect of peculiar 
velocity of galaxies induce 
the anisotropies in the galaxy clustering patterns   
(e.g., \cite{Hamilton:1997zq, Peebles:1980}).   
These are now recognized as a powerful tool to test 
theory of gravity on cosmological scales with a great interest 
(e.g., \cite{Linder:2007nu,Guzzo:2008ac,Yamamoto:2008gr,Song:2008qt,
Song:2010bk}).  
On large scales, the magnitude of redshift distortions 
is simply characterized by the growth-rate parameter $f$, defined as 
$f=d\ln D_+/d\ln a$, where the quantities $D_+$ and $a$ respectively
denote the linear growth factor and the scale factor of the Universe 
\cite{Kaiser:1987qv,1992ApJ385L5H,Hamilton:1997zq}. 
Since the modification of gravity from the general relativity generally 
alters not only the cosmic expansion but also the structure formation,   
the measurement of growth-rate parameter provides a complementary way
to distinguish between various models of gravity.

Notice that distortions of the galaxy 
clustering pattern also arise from apparent mismatch of the 
underlying cosmology when we convert the redshift and angular position 
for each galaxy to the comoving radial and transverse distances. This is 
known as the Alcock-Paczynski effect \cite{Alcock79}, 
and with the baryon acoustic oscillations (BAOs) as a standard ruler, 
it can be utilized for a measurement of both the Hubble 
parameter $H(z)$ and angular diameter distance $D_A(z)$ of distant 
galaxies at redshift $z$ (e.g., \cite{Seo:2003pu,Blake:2003rh,
Glazebrook:2005mb,Shoji:2008xn,Padmanabhan:2008ag}, 
see \cite{Eisenstein05,Cole05,Huetsi06,Percival07,Okumura08,Percival10,Huetsi10,Blake11b,Beutler11} for observational status). 
Hence, a measurement of the anisotropic 
galaxy clustering serves as a dual cosmological probe 
to simultaneously constrain both the cosmic expansion and structure growth 
(e.g., \cite{Taruya11}). 
Planned and ongoing galaxy redshift surveys such as 
the Baryon Oscillation Spectroscopic Survey (BOSS) \cite{Schlegel:2009hj}, 
Hobby-Eberly Telescope Dark Energy Experiment (HETDEX) \cite{Hill:2008mv}, 
Subaru Measurement of Imaging and Redshift equipped with 
Prime Focus Spectrograph (SuMIRe-PFS) \cite{Suto10}, and EUCLID/JDEM 
\cite{Beaulieu:2010qi,Gehrels:2010fn} 
aim at precisely measuring the anisotropic power spectrum and/or two-point 
correlation function in redshift space, from which we can 
simultaneously determine $D_A$, $H$ and $f$ in a very accurate way.

To get tight and robust cosmological constraints, 
an accurate theoretical template for the anisotropic 
power spectrum is highly demanding, 
taking a proper account of various systematic effects. 
In particular, in redshift space, in addition to the 
non-linear gravitational clustering, the clustering statistics 
generally suffer from two competing effects of redshift distortions, 
i.e., enhancement and suppression of the clustering amplitude, referred 
to as the Kaiser and  Finger-of-God effects, respectively 
\cite{Kaiser:1987qv,1983ApJ...267..465D}. While the 
Kaiser effect comes from the coherent motion of matter (or galaxies) and 
the magnitude of this effect is simply described by the growth-rate parameter 
$f$, the Finger-of-God effect is mainly attributed to the virialized random 
motion of the mass residing at a halo, and the significance of this is 
rather sensitive to the properties of the small-scale clustering. 
In the weakly non-linear regime, a tight correlation between 
the velocity and the density fields still remains, and a mixture of Kaiser and 
Finger-of-God effects is expected to be significant. In this sense, 
a careful treatment is needed to model the anisotropic 
power spectrum accurately, otherwise one might derive a biased estimate of the
growth-rate parameter as shown by, e.g., \cite{Okumura11}.

In a previous paper \cite{Taruya10}, 
based on the analytical treatment with perturbation 
theory, we have presented an improved prescription for redshift-space power 
spectrum relevant for the scales of BAOs. 
The model properly takes account of 
both the non-linear gravitational clustering and redshift distortions. 
Contrary to the so-called streaming model 
(e.g., \cite{Peebles:1980,Hatton:1997xs}), which has been 
phenomenologically introduced but frequently used in the literature, 
the new model includes the corrections coming from the non-linear coupling 
between velocity and density fields,  which gives rise to a slight 
increase in the amplitude of the redshift-space power spectrum. Combining 
the improved treatment of perturbation theory to accurately predict non-linear 
clustering \cite{Taruya08,Taruya09}, 
the model reproduces the monopole and quadrupole moments of 
the matter power spectrum measured from N-body simulations quite well, and a 
percent level precision is achieved over the scales of 
interest for BAOs \cite{Taruya10}. 
However, the comparison with N-body simulations 
has been only made with lower multipoles of the matter power spectrum, 
i.e., monopole and quadrupole spectra. 

In this paper, as a natural extension of the previous study, 
we further test the analytical model of redshift distortions. Using a 
large set of N-body simulations, we measure the redshift-space power 
spectrum in two dimensions, characterized as a function of line-of-sight 
wavenumber $k_{||}$ and its normal one $k_\perp$.  
In particular, we examine the halo power spectrum in detail, and 
investigate the extent to which the perturbation theory description 
combining a simple halo bias scheme can describe the 
halo clustering properties in redshift space. We find that 
the measured halo power spectrum in redshift space 
at relatively large scales $k\lesssim0.2\,h\,$Mpc$^{-1}$  
exhibits a significant enhancement in amplitude relative to the real-space 
clustering, which cannot be explained by the streaming 
model. On the other hand, the new model including the corrections can 
describe the enhancement fairly well. The results indicate that 
the halo bias can illuminate the 
non-linear coupling between density and velocity fields, and 
a proper account of the non-linear velocity-density coupling 
seems important in accurately modeling the galaxy/halo power spectrum 
in redshift space.

This paper is organized as follows. We begin with a brief review of
the models based on perturbation theory in redshift space in 
Sec.~\ref{sec:model}. 
Sec.~\ref{sec:nbody} describes the setup of N-body simulations and 
discusses the power spectrum analysis. The results for a detailed 
comparison between models and simulations are presented 
in Sec.~\ref{sec:results}, particularly 
focusing on the halo and matter power spectrum in two 
dimensions.  We discuss about the impact of the non-linear correction terms 
on some typical on-going/planned galaxy redshift surveys in Sec.~\ref{sec:discussion}.
Finally, Sec.~\ref{sec:summary} is devoted to the summary of this paper. 

\section{The models} 
\label{sec:model}

In principle, the clustering statistics in redshift space can be 
mapped from those in real space through the coordinate transformation 
between the positions in real and redshift spaces, ${\bf r}$ and ${\bf s}$
\footnote{Throughout the paper, we work with the distant-observer 
approximation, and neglect the angular dependence of the 
line-of-sight direction, relevant for the high-redshift galaxy surveys.}: 
\be
{\bf s} = {\bf r} + \frac{1+z}{H(z)}\,v_z({\bf r})\,\widehat{\bf z},
\label{eq:def_s-space}
\ee
where the quantity $H$ is the Hubble parameter 
at redshift $z$, the unit vector $\hat{z}$ indicates the line-of-sight 
direction, and the quantity $v_z$ represents the line-of-sight 
component of the peculiar velocity field. Although the expression 
(\ref{eq:def_s-space}) is very simple,  
the statistical relationship between the real- and redshift-space 
quantities is rather complicated and is actually difficult to treat 
without approximations. One reason for the difficulty comes from the 
anisotropies induced by the velocity field which is very sensitive to the 
small-scale structure. The other important aspect of the redshift-space 
clustering is the coupling between velocity and density fields, which 
produces an apparent structure growth by the coherent motion.

Qualitatively, the clustering properties in redshift space are explained by 
the Kaiser and Finger-of-God effects. The so-called streaming model 
is a phenomenological model that accounts for these two effects separately 
\cite{Peebles:1980,Hatton:1997xs}. 
Here, among various streaming models,  
we consider the following form of the redshift-space power spectrum 
(e.g., \cite{Scoccimarro04,Percival:2008sh}): 
\begin{align}
&P(k,\mu) = D_{\rm f}(k\mu\,f\,\sigma_{\rm v})\,
\nonumber\\
&\qquad\qquad\times\,\,\left[P_{\delta\delta}(k)+2\,f\mu^2P_{\delta\theta}(k)
+f^2\mu^4P_{\theta\theta}(k)\right],
\label{eq:SCO}
\end{align}
where $f$ is the growth rate parameter, 
the quantity $\sigma_{\rm v}$ is the one-dimensional velocity dispersion,  
and $\mu$ is the directional cosine of the angle between 
line-of-sight $\hat{\bf z}$ and the Fourier mode $\bfk$. 
In the above, while the function $D_{\rm f}$ represents 
a damping function mimicking the Finger-of-God effect, 
the term in the bracket indicates an improved prescription of the 
Kaiser effect which takes account of the 
nonlinear gravitational evolution \cite{Scoccimarro04}. 
The spectra $P_{\delta\delta}$, $P_{\theta\theta}$, and $P_{\delta\theta}$ 
respectively 
denote the auto-power spectra of density and velocity divergence, and their 
cross-power spectrum, with the velocity divergence $\theta$ defined by 
$\theta\equiv-(1+z)/(Hf)\nabla{\bf v}$. The explicit functional form of the 
damping function will be specified later.

Notice that the streaming models have been originally introduced and 
frequently used in the literature to explain the observed power spectrum 
on small scales.  
On large scales of our interest, non-linearity of the gravitational clustering 
is rather mild, but a tight correlation between density and 
velocity fields still remains. 
In previous 
paper \cite{Taruya10}, starting with the rigorous expression of 
redshift-space power spectrum, we partially applied a low-$k$ expansion, 
and found that the model (\ref{eq:SCO}) misses 
some important terms. The missing terms naturally 
arise from the next-to-leading order corrections of the low-$k$ expansion, 
and they represent the nonlinear coupling between velocity and density 
fields. The model is given by 
\begin{align}
&P(k,\mu) = D_{\rm f}(k\,\mu\,f\sigma_{\rm v})\,
\nonumber\\
&\quad\times\,\Bigl[P_{\delta\delta}(k)+2\,f\mu^2P_{\delta\theta}(k)
+f^2\mu^4P_{\theta\theta}(k)\Bigr.
\nonumber\\
&\qquad\qquad\qquad\qquad\qquad\quad
\Bigl.+A(k,\mu;f)+B(k,\mu;f)\Bigr]. 
\label{eq:TNS}
\end{align}
The newly derived terms $A$ and $B$ are written as 
\begin{align}
&A(k,\mu;f)= (k\mu\,f)\,\int \frac{d^3\bfp}{(2\pi)^3} \,\,\frac{p_z}{p^2}
\nonumber\\
&\qquad\quad\times
\left\{B_\sigma(\bfp,\bfk-\bfp,-\bfk)-B_\sigma(\bfp,\bfk,-\bfk-\bfp)\right\},
\label{eq:A_term}
\\
&B(k,\mu;f)= (k\mu\,f)^2\int \frac{d^3\bfp}{(2\pi)^3} F(\bfp)F(\bfk-\bfp)\,\,;
\label{eq:B_term}
\\
&\qquad\quad F(\bfp)=\frac{p_z}{p^2}
\left\{ P_{\delta\theta}(p)+f\,\frac{p_z^2}{p^2}\,P_{\theta\theta}(p)\,\right\},
\nonumber
\end{align}
where the function $B_\sigma$ is the cross bispectra defined by 
\begin{align}
&\left\langle \theta(\bfk_1)
\left\{\delta(\bfk_2)+f\,\frac{k_{2z}^2}{k_2^2}\theta(\bfk_2)\right\}
\left\{\delta(\bfk_3)+f\,\frac{k_{3z}^2}{k_3^2}\theta(\bfk_3)\right\}
\right\rangle
\nonumber\\
&\quad\qquad
=(2\pi)^3\delta_D(\bfk_1+\bfk_2+\bfk_3)\,B_\sigma(\bfk_1,\bfk_2,\bfk_3).
\label{eq:def_B_sigma}
\end{align}
Within the range of the validity of the low-$k$ expansion,  
these corrections should be small, and we apply the standard PT treatment 
to calculate the $A$ and $B$ terms. To compute 
the corrections, we use the expressions (A3) and (B4) presented in 
Ref.~\cite{Taruya10}, which are suited for a fast numerical calculation.

The previous study \cite{Taruya10} reveals that the corrections slightly 
enhance the power spectrum amplitude over the scales of BAOs, and 
the model (\ref{eq:TNS}) excellently reproduces 
the monopole and quadrupole spectra obtained from N-body simulations. 
Note, however, that the aforementioned models of redshift distortions are 
those for the matter distribution, and the previous study restricted 
the analysis to the case of the matter power spectrum. 
In this paper, adopting a simple linear bias relation without no 
velocity bias in real space,  i.e., $\delta_{\rm h}=b\,\delta_{\rm m}$, 
we further test the 
model (\ref{eq:TNS}) against the halo clustering in redshift space. 
Then, the model of halo power spectrum becomes
\begin{align}
&P_{\rm h}(k,\mu) = D_{\rm f}(k\,\mu\,f\sigma_{\rm v})\,
\nonumber\\
&\quad\times\,b^2\Bigl[P_{\delta\delta}(k)+2\,\beta\,\mu^2P_{\delta\theta}(k)
+\beta^2\,\mu^4P_{\theta\theta}(k)\Bigr.
\nonumber\\
&\qquad\qquad\qquad\qquad\quad
\Bigl.+b\,A(k,\mu;\beta)+b^2\,B(k,\mu;\beta)\Bigr] 
\label{eq:TNS_halo}
\end{align}
with the quantity $\beta$ defined by $\beta=f/b$. 
We use the improved PT treatment developed by Refs.~\cite{Taruya08,Taruya09} 
to compute $P_{\delta\delta}$, $P_{\theta\theta}$ and $P_{\delta\theta}$. 
We will check the validity and accuracy of the model prescription 
with and without the corrections $A$ and $B$ in detail for the halo 
clustering. In what follows, we specifically call Eq.~(\ref{eq:TNS_halo}) 
including the corrections the TNS model, and discriminate it from 
Eq.~(\ref{eq:TNS_halo}) neglecting the corrections as the streaming model. 
As for the damping function $D_{\rm f}$, we adopt the Gaussian and 
the Lorentzian forms often used in the literature 
(e.g., \cite{Peacock:1993xg,Park:1994fa,Ballinger:1996cd,
Magira:1999bn}), and compare between the results of these predictions: 
\be
D_{\rm f}(x) = 
\left\{
\begin{array}{l}
\displaystyle\exp(-x^2),\label{eq:gauss}\\
\displaystyle\frac{1}{(1+x^2/2)^2}.\label{eq:lorentz}
\end{array}
\right.
\ee
Since the damping function mainly alters the global shape of the 
power spectrum and our primary focus is to model the acoustic 
feature precisely, we will treat the velocity dispersion $\sigma_{\rm v}$ 
in the damping function as a free parameter, and determine 
it by fitting the prediction to the N-body simulations.

Finally, note that in the case of the TNS model, 
the expression (\ref{eq:TNS_halo}) is 
valid only when the linear bias parameter $b$ is 
scale-independent. Nevertheless, in later analysis, 
we allow to incorporate the scale-dependence of the bias into the model 
(\ref{eq:TNS_halo}), and compare it with the halo power spectrum in 
redshift space. Strictly speaking, the 
scale-dependence of the bias changes the structure of the 
integral kernel in Eqs.~(\ref{eq:A_term}) and (\ref{eq:B_term}), and 
we cannot use the formulas for the corrections $A$ and $B$ presented 
in Ref.~\cite{Taruya10}, which have been derived for the matter 
power spectrum. As shown in Sec.~\ref{subsec:r-space}, however, 
the scale-dependence of the halo bias calibrated from 
the N-body simulations turned out to be very weak over the scales of 
our interest. Thus, our treatment of the model (\ref{eq:TNS_halo}) with 
scale-dependent bias would be validated, and does not change 
the final conclusion.

\section{Analysis}
\label{sec:nbody}

\subsection{N-body simulations}
\label{subsec:simulations}

To assess the validity and accuracy of the analytic models against the 
redshift-space halo clustering, we run a set of N-body simulations and 
identified halos. All the simulations are performed with a publicly 
available tree PM code {\tt GADGET2}
\cite{Springel05}. We adopt $N=1,280^3$ particles in boxes with a side 
length of $1,144.72\,h^{-1}$ Mpc [i.e., the volume is $1.5 \,(h^{-1}$Gpc$)^3$], 
and set the softening length to
$5\%$ of the inter-particle distance. We generate 
the initial conditions at $z=99$ by second-order Lagrangian
perturbation theory (e.g., \cite{Scoccimarro98,Crocce06}) starting from 
particles put on the regular lattices. 
The matter transfer function is calculated by a publicly available 
Boltzmann solver {\tt CAMB} \cite{Lewis00},
assuming the best-fit $\Lambda$CDM cosmological model determined by 
the five-year WMAP \cite{Komatsu09}.
The output data of $15$ independent realizations are stored at $z=0.35$. 
Table~\ref{tab:sims} summarizes the settings of the simulations.

\begin{table*}[!ht]
\caption{Summary of simulation parameters.}
\begin{ruledtabular}
\label{tab:sims}
\begin{tabular}{c|c|c|c|c|c|c|c|c|c|c|c}
 \# of runs & \# of particles & box size & softening & mass/particle & $z_{\rm in}$ & $z_{\rm out}$ &
 $\Omegam$ & $\Omegab/\Omegam$ & $h$ & $n_s$ & $\sigma_8$\\
\hline
\hline
$15$ & $1,280^3 $ & $1,144.72h^{-1}$Mpc & $44.72h^{-1}$kpc & $5.54\times10^{10}h^{-1}M_\odot$ & $99$ & $0.35$
& $0.279$ & $0.165$ & $0.701$ & $0.96$ & $0.817$
\end{tabular}
\end{ruledtabular}
\end{table*}

\subsection{Halo catalogs}
\label{subsec:halo}

Given the data set of dark matter clustering, 
we identify the halos by a Friends-of-Friends finder with linking length 
$0.2$ times the mean inter-particle distance. For each realization, 
we construct nine halo catalogs with
different minimum and maximum masses. Allowing a slight overlap of the 
mass range, each mass bin is determined so that 
the signal-to-noise ratio of halo power spectrum
measured at $k=0.0975h$Mpc$^{-1}$ becomes 
roughly comparable among the nine halo catalogs 
[see Eq.~(\ref{eq:def_SN}) for the definition of signal-to-noise ratio]. 
Detailed properties of the nine halo catalogs, including the 
mass range and the number density of halos, are summarized in 
Table~\ref{tab:halo}. In what follows, we especially call 
the least and most massive halo catalogs (bin $1$ and $9$) 
as {\tt light} and {\tt heavy}, and discuss their 
clustering properties in detail. 
Note that the {\tt heavy} catalog has roughly the same 
values of the bias parameter, number density, and survey volume as 
observed in the luminous red galaxies (LRG) \cite{Eisenstein01} of 
the seventh data release of Sloan Digital Sky Survey (SDSS DR7) 
\cite{Abazajian09}.

\begin{table*}[!ht]
\begin{ruledtabular}
\caption{
Summary of the halo catalogs. The minimum, maximum and mean mass ($M_{\rm min}$, $M_{\rm max}$
and $\overline{M}_{\rm h}$) are in units of $h^{-1}M_\odot$, while the halo number density ($n_h$) is in
$h^3$Mpc$^{-3}$. The bias parameter, $b_0$, is defined in Eq.~(\ref{eq:Q}). See Sec.~\ref{subsec:r-space} 
for more detail.
}
\label{tab:halo}
\begin{tabular}{c||c|c|c|c|c|c|c|c|c}
 Sample & bin 1 ({\tt light}) & bin 2 & bin 3 & bin 4 & bin 5 & bin 6 & bin 7 & bin 8 & bin 9 ({\tt heavy})\\
 \hline\hline
 $M_{\rm min}$ & $1.77\times10^{12}$ & $2.49\times10^{12}$ & $3.54\times10^{12}$ & $4.98\times10^{12}$ & $7.09\times10^{12}$ & 
 $1.00\times10^{13}$ & $1.42\times10^{13}$ & $2.01\times10^{13}$ & $2.84\times10^{13}$ \\
 \hline
 $M_{\rm max}$ & $5.54\times10^{12}$ & $1.02\times10^{13}$ & $1.74\times10^{13}$ & $2.66\times10^{13}$ & $4.04\times10^{13}$ &
 $6.76\times10^{13}$ & $1.19\times10^{14}$ & $2.08\times10^{14}$ & - \\
 \hline
   $\overline{M}_{\rm h}$ & $2.96\times10^{12}$ & $4.65\times10^{12}$ & $7.08\times10^{12}$ & $9.37\times10^{12}$ & $1.47\times10^{13}$ &
 $2.18\times10^{13}$ & $3.21\times10^{13}$ & $4.63\times10^{13}$ & $7.03\times10^{13}$ \\
 \hline
$n_h$ & $1.57\times10^{-3}$ & $1.26\times10^{-3}$ & $9.46\times10^{-4}$ & $6.87\times10^{-4}$ & $4.87\times10^{-4}$ & 
 $3.47\times10^{-4}$ & $2.43\times10^{-4}$ & $1.64\times10^{-4}$ & $1.09\times10^{-4}$ \\
  \hline
 $b_0$ & $1.08$ & $1.16$ & $1.25$ & $1.35$ & $1.47$ & $1.62$ & $1.80$ & $1.99$ & $2.26$ 
\end{tabular}
\end{ruledtabular}
\end{table*}

\subsection{Power spectrum}
\label{subsec:analysis}

We measure the dark matter and halo power spectra from the N-body 
simulations using the standard method based on the fast Fourier 
transform. To be specific, we first assign the dark matter or halo 
particles onto a $1,024^3$ grid by cloud-in-cells (CIC) interpolation scheme 
\cite{Hockney81} to obtain a density field on a lattice. We then transform 
it to Fourier space and divide it by the CIC kernel in order to eliminate
the window effect.  
Finally, we take the average of the power over the modes in each 
$k$ bin in real space, 
and in $(k, \mu)$ bin in redshift space. The size of each Fourier bin 
is set to $\Delta k=0.005h$Mpc$^{-1}$ and $\Delta \mu=0.05$ for the 
wavenumber and the directional cosine, respectively.

Our primary focus is to test the analytic models of redshift 
distortions against the halo clustering in N-body simulations. To do this, 
we need to incorporate the halo bias properties characterized by the 
function $b(k)$ into the model prediction 
(\ref{eq:TNS_halo}). While, in practice, the quantity 
$b(k)$ should be parametrized by a simple function, and 
must be determined by fitting the model prediction to the observed power 
spectrum, in order to make a transparent test, 
we directly measure the halo bias parameter $b(k)$ from the N-body 
simulation, and use it to compute the analytic models of redshift-space 
power spectrum. Note that this treatment still includes several 
non-trivial aspects. Indeed, the linear bias prescription seems 
rather simplistic, and it might not capture the real clustering nature of 
halos over the BAO scales. Further, in redshift space, the presence of 
correction terms in the TNS model implies that even the simple linear bias 
can lead to a non-trivial modulation of clustering amplitude, which 
might drastically alter the acoustic structure in redshift space.

Let us discuss how to characterize and measure the halo bias properties 
from N-body simulations. According to the treatment by Ref.~\cite{Seljak09}, 
we decompose the halo density contrast defined in real space into two pieces: 
\be
\deltah(\bfk) = b(k)\deltam(\bfk)+\epsilon(\bfk).
\label{eq:def_bias}
\ee
Here, the quantity $\epsilon$, which satisfies 
$\langle\deltam\epsilon\rangle=0$, can be regarded as the residual 
noise contribution, and is  
related to the stochasticity of the halo sampling process. 
Notice that the above decomposition does not rely on 
the assumption of linear bias. The other additional contributions,  
which cannot be simply written as the term linearly proportional to 
$\deltam$, are all incorporated into the second term. 
Given the auto and cross power spectra of 
matter and halos measured in real space, $\Pm$, $\Ph$ and $\Phm$,  
the decomposition of the signal linearly proportional to the matter 
density field, and noise can be uniquely done, and the halo bias parameter
can be measured in the following way: 
\be
b(k) &=& \Phm(k)/\Pm(k),
\label{eq:b_from_pk}
\\
P_\epsilon(k) &=& \Ph(k)-b^2(k)\Pm(k), 
\label{eq:eps_from_pk}
\ee
where the quantity $P_\epsilon$ is the residual noise power spectrum 
defined by $\langle\epsilon(\bfk)\epsilon(\bfk')\rangle
=(2\pi)^3\delta_D(\bfk+\bfk')P_\epsilon(k)$. In deriving 
the above relations, 
we have used the property $\langle\deltam\epsilon\rangle=0$. 
Note that the halo subsamples
listed in Table~\ref{tab:halo} are constructed so as to 
take the same signal-to-noise ratio 
among the nine catalogs. The signal-to-noise ratio is defined by
\be
\left(\frac{\rm S}{\rm N}\right)_k=\frac{b^2(k)\Pm(k)}{P_\epsilon(k)}. 
\label{eq:def_SN}
\ee
At $k=0.0975\,h$Mpc$^{-1}$, the signal-to-noise ratio roughly reaches
$({\rm S/N})_k\simeq4$ for all of our subsamples.

In testing the analytic models of redshift-space power spectrum, 
we subtract the residual noise contribution from the measured halo 
spectrum, and then compare the results with analytic prediction 
in the $(k,\mu)$ space. Note that apart from the 
bias $b(k)$, the models (\ref{eq:SCO}) and (\ref{eq:TNS}) 
[or (\ref{eq:TNS_halo})] involve a single parameter $\sigma_{\rm v}$,  
which has to be determined by fitting the model power spectrum to 
the measured power spectrum. To derive the best-fit value of 
$\sigma_{\rm v}$, we assume the Gaussianity, 
and use the statistical error of the redshift-space 
power spectrum given by \cite{Feldman94}
\be
\left[\Delta P(k,\mu)\right]^2 &=& 
\frac{1}{N_{\rm mode}}\left[b^2(k)\Pm(k,\mu)+P_\epsilon(k)\right]^2,
\label{eq:pk_error}
\ee
where $N_{\rm mode}$ denotes the total number of 
independent Fourier modes which fall into the $(k, \mu)$ bin over the 
$15$ realizations. Note that in estimating the best-fit $\sigma_{\rm v}$ 
for dark matter, we simply set $b(k)=1$ and $P_\epsilon=0$.

Finally, in discussing the goodness of fit between 
the streaming and TNS models [i.e., Eq.(\ref{eq:TNS_halo}) with and without 
corrections], it is convenient to introduce the reduced chi-squared statistic:  
\be
\chi_{\rm red}^2=\frac{1}{\nu}\sum_{i,j} 
\frac{\left[P_{\rm N\mbox{-}body}(k_i,\mu_j)-P_{\rm model}(k_i,\mu_j)\right]^2}
{\left[\Delta P(k_i,\mu_j)\right]^2},
\label{eq:chisquared}
\ee
where the power spectra $P_{\rm N\mbox{-}body}$ and $P_{\rm model}$ are 
respectively 
obtained from the N-body simulations and the analytic models with the best-fit 
value of $\sigma_{\rm v}$. 
The quantity $\nu$ is the number of degrees of freedom, and depending on 
the range of fitting, wet set $\nu=204$, $364$, $524$, and $684$
for the maximum wavenumber used to fit, 
$k_{\rm max}=0.08$, $0.12$, $0.16$, and $0.2\,h$Mpc$^{-1}$, respectively.

\section{Results}
\label{sec:results}

In this section, the results of the power spectrum measurement are presented,
and a detailed comparison between N-body simulations with analytic models 
is made. We first address the real-space clustering for dark matter and 
halos, and measure the halo bias in Sec.~\ref{subsec:r-space}. 
We then move to the discussion on the redshift-space power spectrum, 
and the clustering anisotropies caused by redshift distortions are shown 
in two-dimensional plane in Sec.~\ref{subsec:2d}. 
In Sec.~\ref{subsec:multipole}, applying the 
multipole expansion to the anisotropic power spectra, 
the lowest three multipole spectra, 
i.e., monopole, quadrupole, and hexadecapole spectra, are quantified 
and compared with analytic models. 
While we mainly analyze the dark matter, and {\tt light} and {\tt heavy} halo 
catalogs, we also examine the other halo catalogs in 
Sec.~\ref{subsec:accuracy}, and study how well the 
analytic models can reproduce the N-body results when we vary 
the halo mass and/or fitting range.

\subsection{Real-space clustering}
\label{subsec:r-space}

\begin{figure}[b] 
   \centering
   \includegraphics[height=8.4cm]{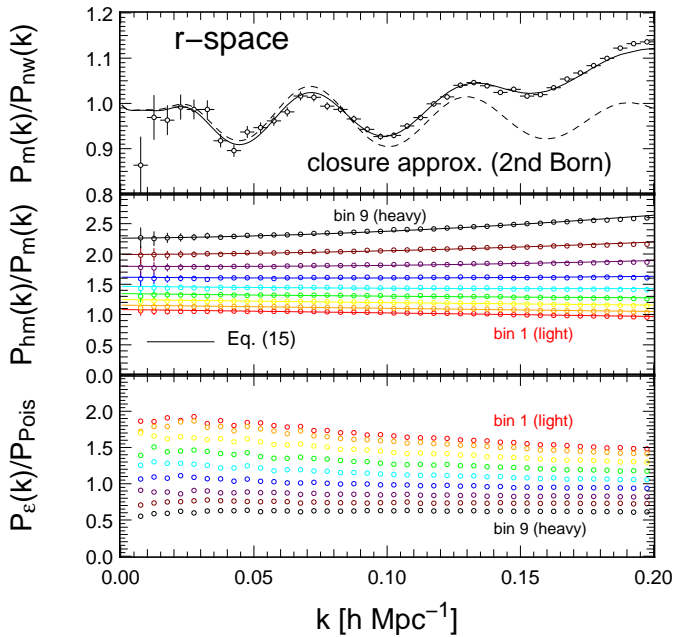} 
   \caption{Ratios of the real-space power spectra for dark matter and halos. 
   {\it Top}: matter power spectrum
   divided by the no-wiggle linear power spectrum in \cite{EH98}. 
   The symbols are N-body data, 
   while the lines are analytical predictions (solid: Closure Approximation up to the 2nd order of the 
   Born approximation \cite{Taruya08}, dashed: linear theory). 
   {\it Middle}: bias parameter defined by the ratio of the halo-matter cross power spectrum to the matter auto power 
   spectrum [symbols: N-body data, lines: fitted results of Eq.~(\ref{eq:Q})]. The results for the 
   halo samples of bin $1$, $2$, ..., and $9$ are shown from bottom to top. {\it Bottom}: 
   residual noise power spectrum, $P_\epsilon(k)$, divided by the Poisson noise 
   (bin $1$, $2$, ..., and bin $9$ from top to bottom).}
   \label{fig:r-space}
\end{figure}

Since the validity and precision of analytic models given in 
Eq.~(\ref{eq:TNS_halo}) heavily rely on the 
accuracy of the prediction in real space, it is important to first 
check the PT treatment, and to compare the PT prediction with 
N-body simulations. Top panel of 
Fig.~\ref{fig:r-space} shows the ratio of the matter power spectrum 
to the smooth reference spectrum, $\Pm(k)/P_{\rm nw}(k)$, where 
the reference spectrum $P_{\rm nw}(k)$ is calculated from the 
linear transfer function of no-wiggle approximation by Ref.~\cite{EH98}. 
While the dashed line represents the linear theory prediction, the 
solid line is the result based on the improved PT calculation by 
Refs.~\cite{Taruya08,Taruya09}, including the next-to-leading order 
non-linear corrections up to the second-order Born approximation. 
At $z=0.35$, the development of non-linear gravitational clustering is 
bit significant, and even the BAO scales at $k\lesssim0.2\,h$Mpc$^{-1}$ exhibit 
a prominent enhancement in the power spectrum amplitude. Nevertheless, the 
improved PT prediction reproduces the N-body result fairly well, and 
the agreement between simulations and PT prediction is mostly within the 
errorbars estimated from the $15$ realizations of N-body simulations. 
This indicates that the improved PT by 
Refs.~\cite{Taruya08,Taruya09} can be reliably applied to the 
modeling of redshift-space power spectrum at 
$k\lesssim0.2\,h$Mpc$^{-1}$, where the acoustic signature is still visible.

In addition to the PT treatment, 
the halo clustering bias measured in real space is also an 
important building block to precisely model the redshift-space 
power spectrum of halos. Middle and bottom panels of Fig.~\ref{fig:r-space} 
quantify the halo bias properties based on the expressions 
(\ref{eq:b_from_pk}) and (\ref{eq:eps_from_pk}). Middle panel plots 
the halo bias parameter $b(k)$, while bottom panel shows the 
residual noise spectrum divided by the Poisson noise 
$P_{\rm Pois}\equiv 1/n_h$, where $n_h$ is the mean number density of 
halos in each subsample (see Table~\ref{tab:halo}). 
The halo mass dependence is clearly seen not 
only in the halo bias but also in the residual noise spectrum. 
As increasing the halo mass, while the bias parameter increases, the 
residual noise relative to the Poisson error monotonically 
decreases and manifests a non-Poissonian feature, 
i.e., $P_\epsilon/P_{\rm Pois}\ne1$. 
Although the mass dependence of the bias parameter is qualitatively 
and even quantitatively well-understood based on the halo model 
prescription (see \cite{Cooray02} for a review), 
the non-Poissonian feature of the
residual noise indicates that the origin of the residual noise might not be 
simply explained by the halo sampling process. 
Recently, an attempt to reproduce the non-Poissonian feature of the
residual noise has been made, and the mass dependence of the residual 
noise is shown to be explained by the halo model 
\cite{Hamaus10}. Though we do not discuss at all a quantitative 
aspect of this, the non-Poissonian feature of the residual noise 
potentially affects the power spectrum estimation, and the understanding 
of it is practically important to extract a pure clustering signal.

Apart from the origin of the residual noise, it is worth noting that 
the scale-dependence of the halo bias as well as the noise power spectrum 
is quite moderate over the scales of our interest. 
This partly validates our treatment in computing the 
power spectrum from the TNS model (\ref{eq:TNS_halo}) 
(see Sec.~\ref{sec:model}).   
The weak scale-dependence of the halo bias can be well-described by a 
simple fitting function used in the literature \cite{Cole05}:
\be
b^2(k) = b_0^2\,\frac{1+Qk^2}{1+Ak}. 
\label{eq:Q}
\ee
As a reference, the fitted results of Eq.~(\ref{eq:Q}) are shown in 
solid lines in Fig.~\ref{fig:r-space}, and 
the best-fit values of the parameter $b_0$,  
which represents the clustering bias in the large-scale limit, 
are listed in 
Table~\ref{tab:halo} \footnote{Note that in fitting Eq.~(\ref{eq:Q}) to 
the measured bias parameter, we allow to vary $A$ as well as 
$b_0$ and $Q$, although the model (\ref{eq:Q}) has been originally 
introduced adopting the fixed value, $A=1.4$. }. The resultant 
best-fit value $b_0$ of the {\tt heavy} sample is quite close to the 
observed value of the luminous red galaxies in the SDSS DR7.

\subsection{Redshift-space clustering in two dimensions}
\label{subsec:2d}

We are in position to discuss the clustering properties 
of dark matter and halos in redshift space, and to examine the validity of 
the analytic model prescription. In this subsection, we particularly 
focus on the redshift-space power spectrum in two dimensions, and 
compare the model predictions with N-body simulations.

\begin{figure*}[!t] 
   \centering
   \includegraphics[height=5.4cm]{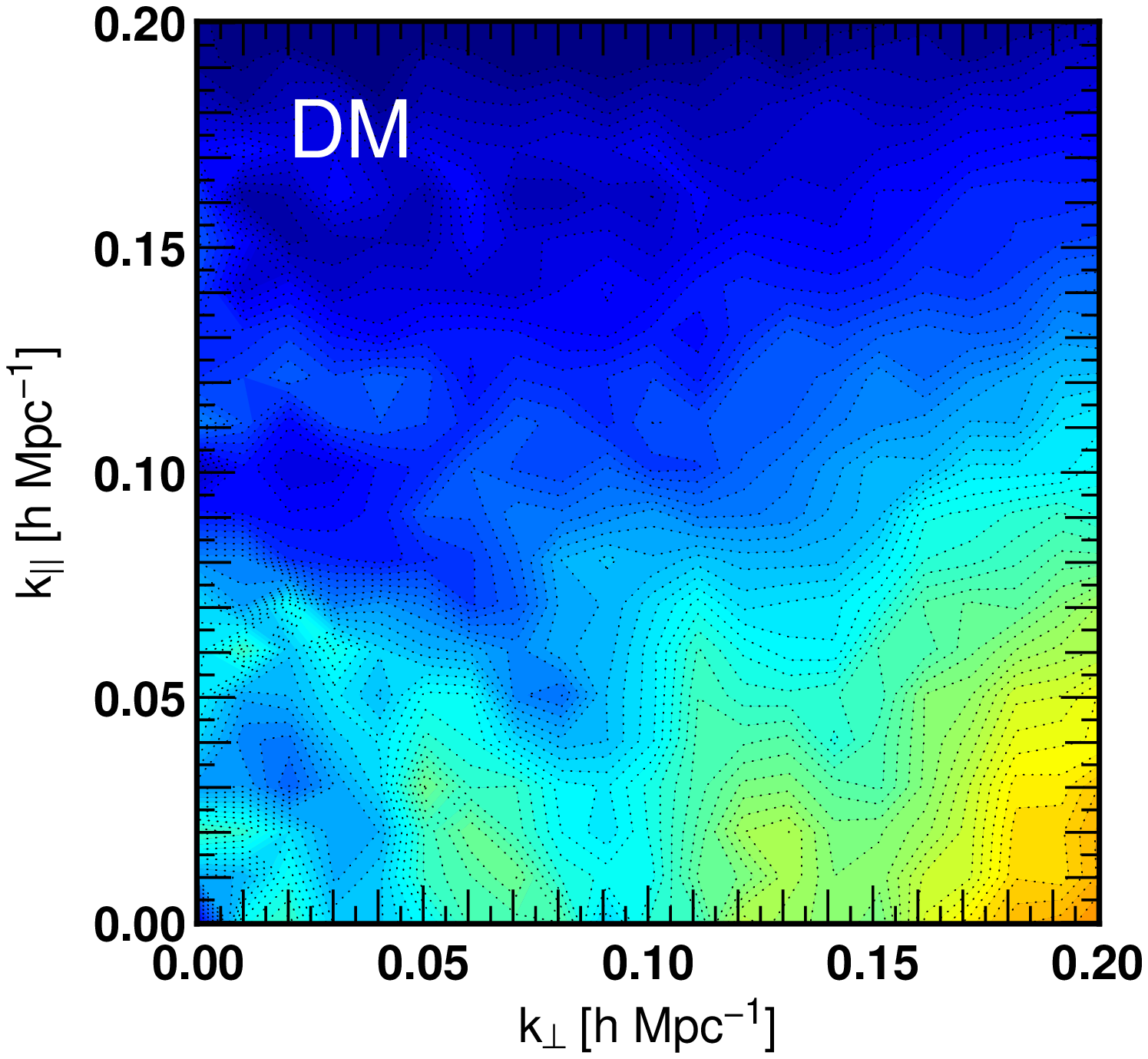} 
   \includegraphics[height=5.4cm]{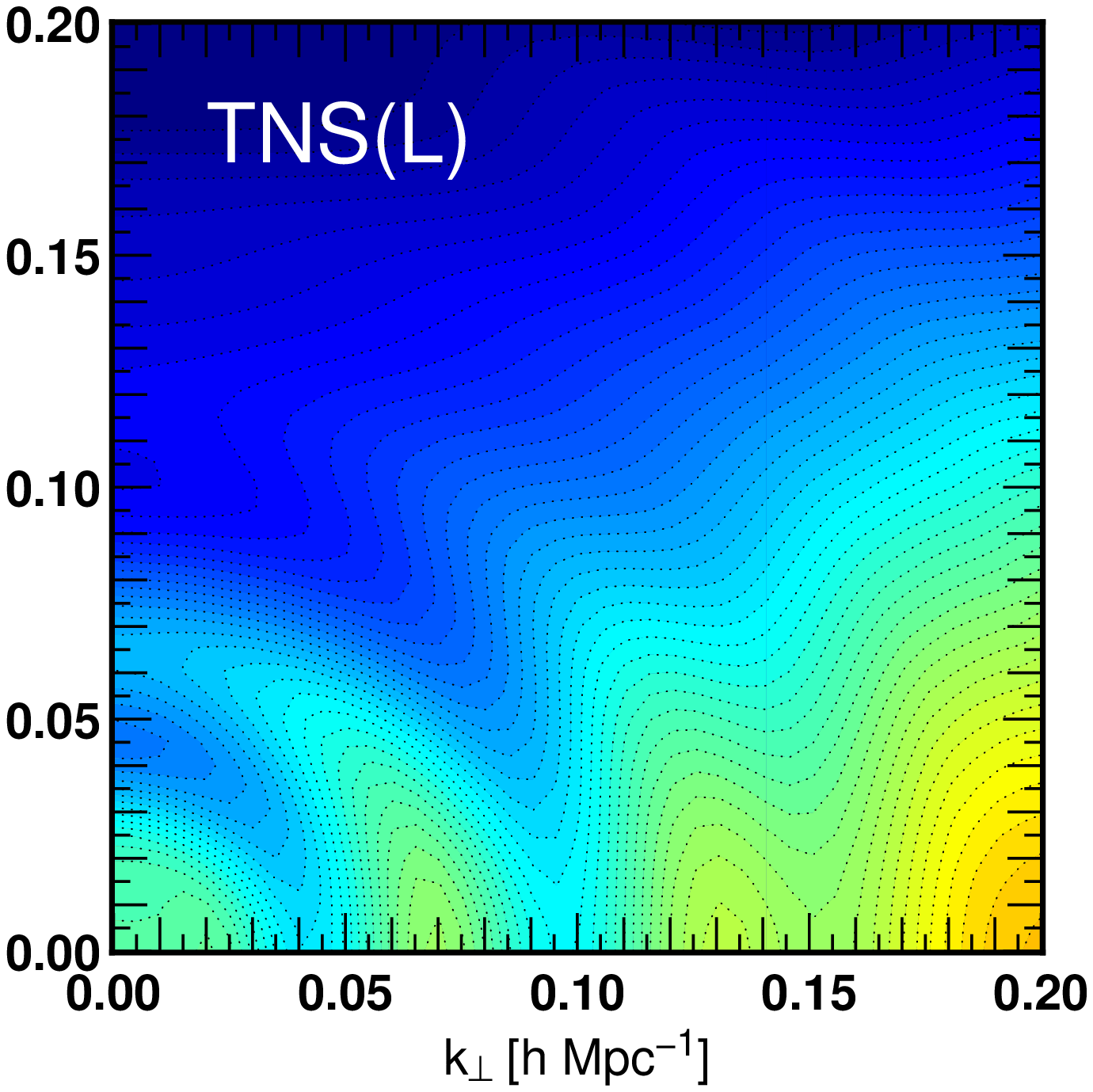} 
   \includegraphics[height=5.4cm]{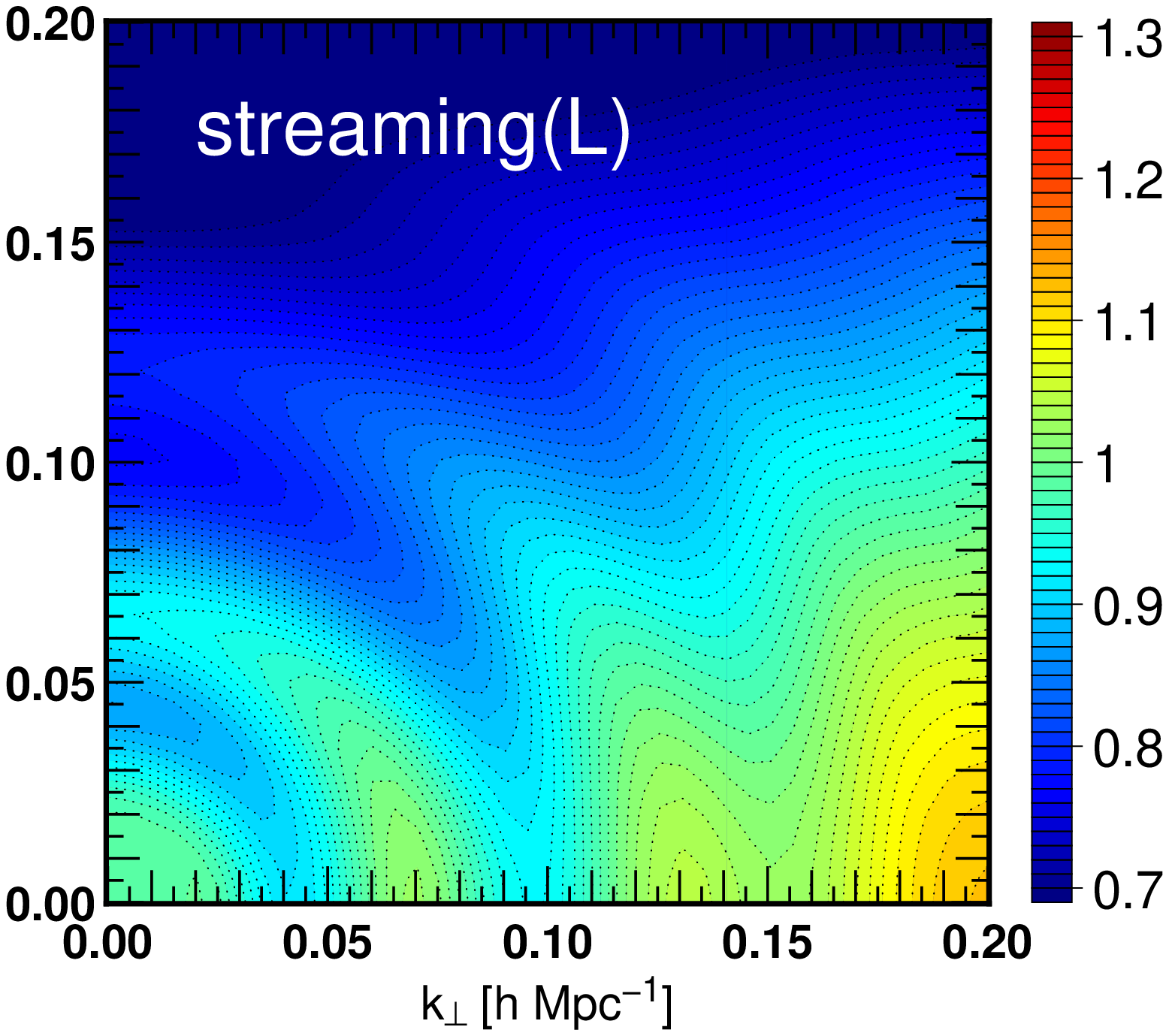} 
   \includegraphics[height=5.4cm]{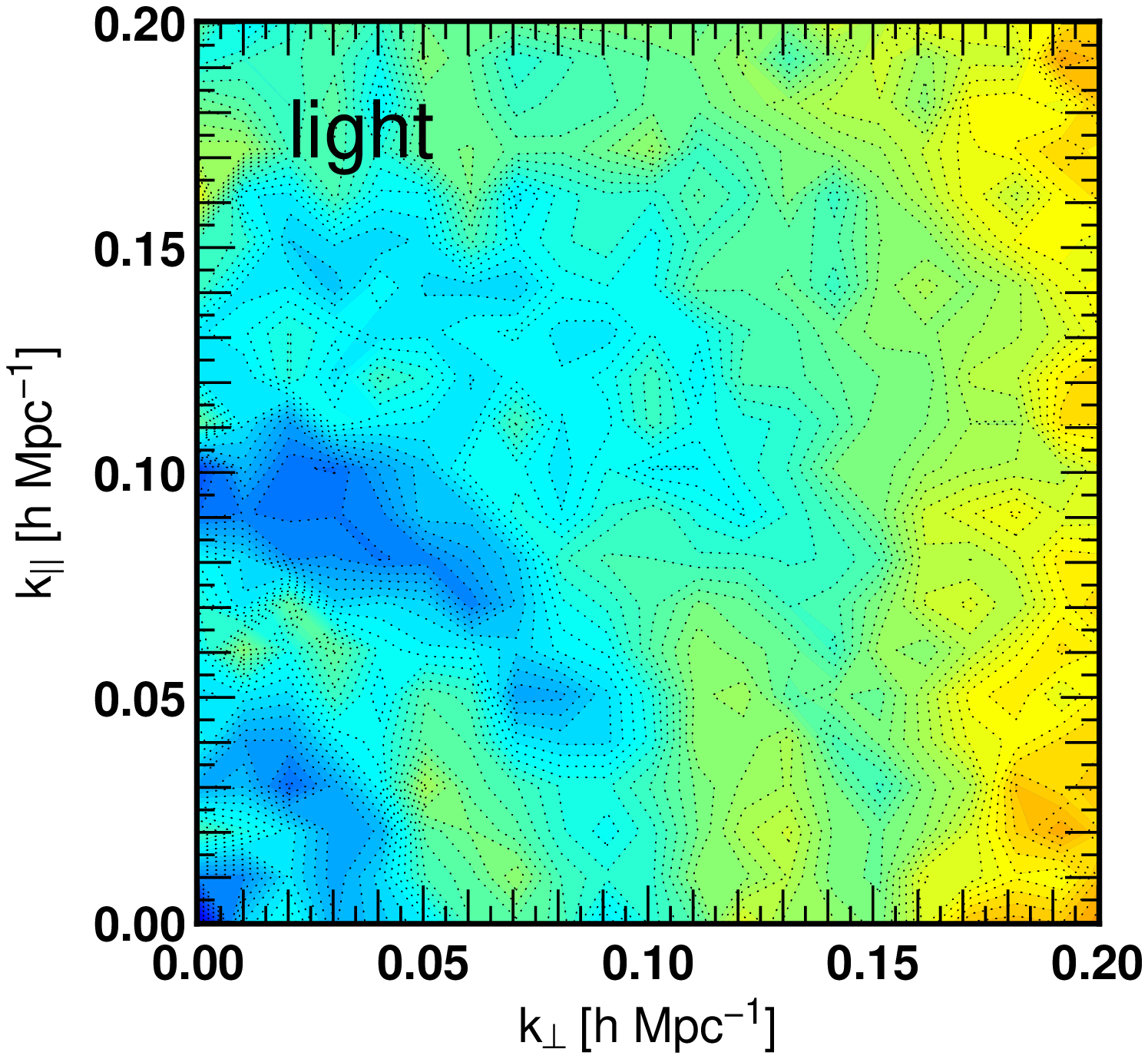} 
   \includegraphics[height=5.4cm]{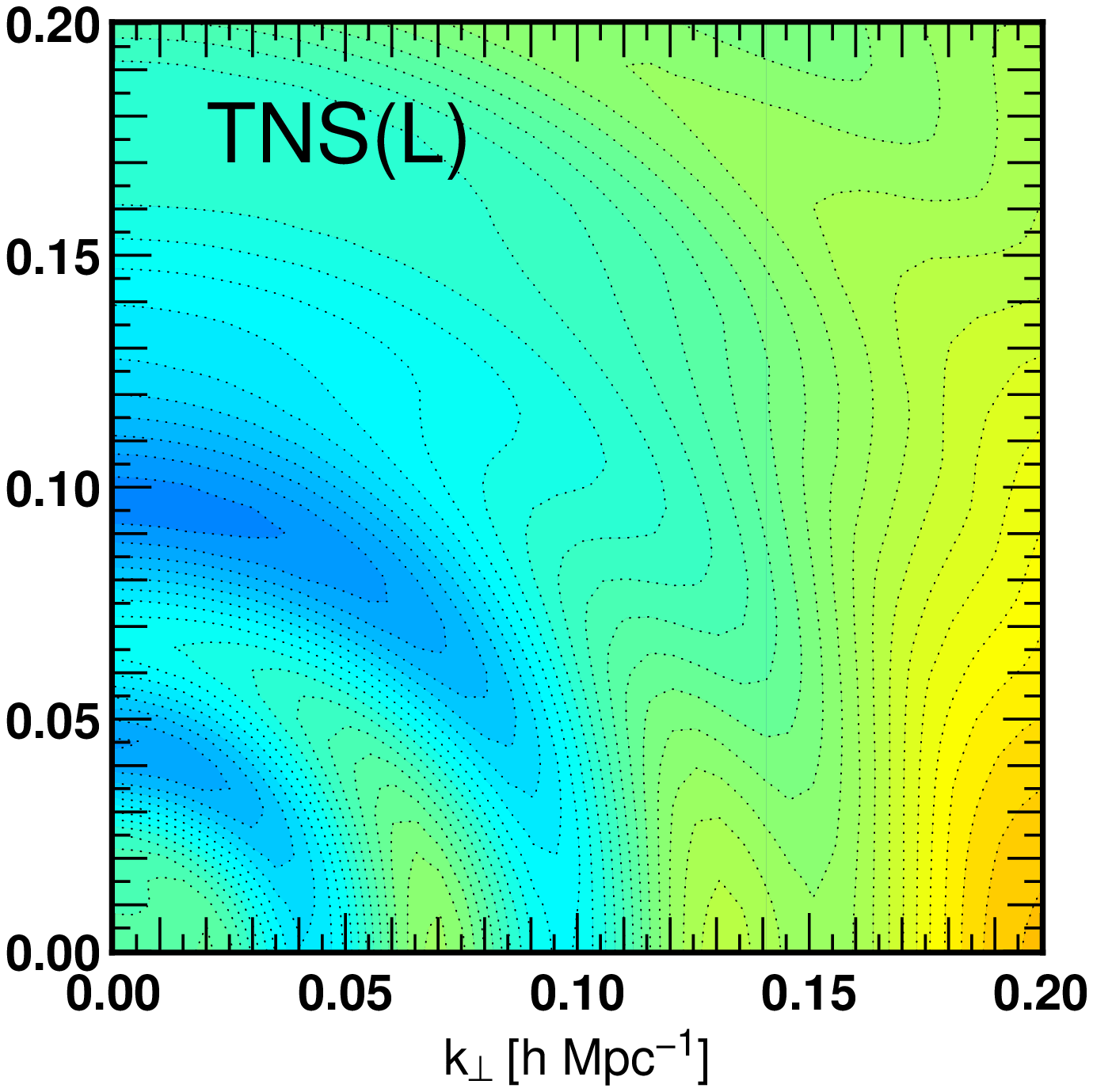} 
   \includegraphics[height=5.4cm]{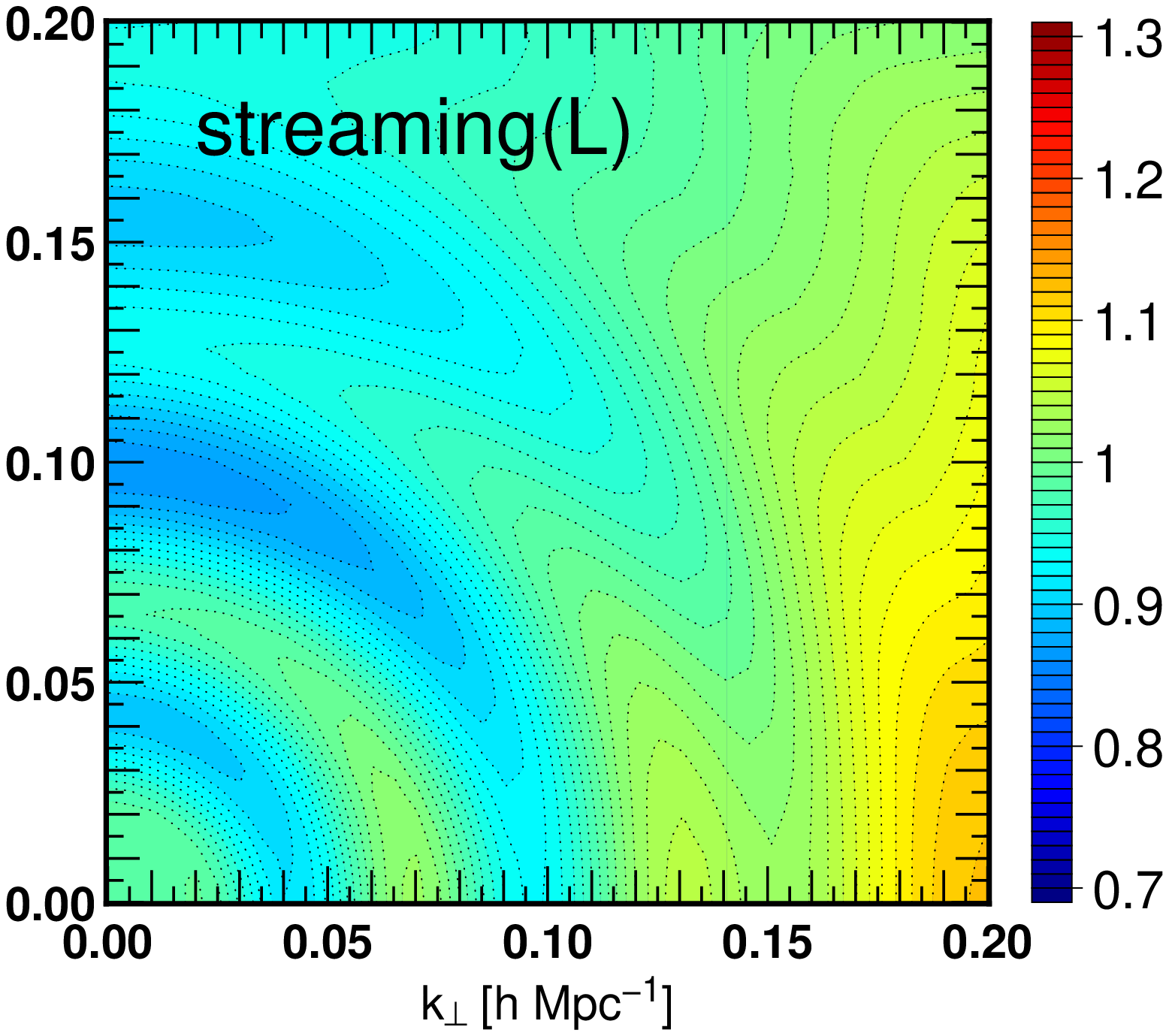} 
   \includegraphics[height=5.4cm]{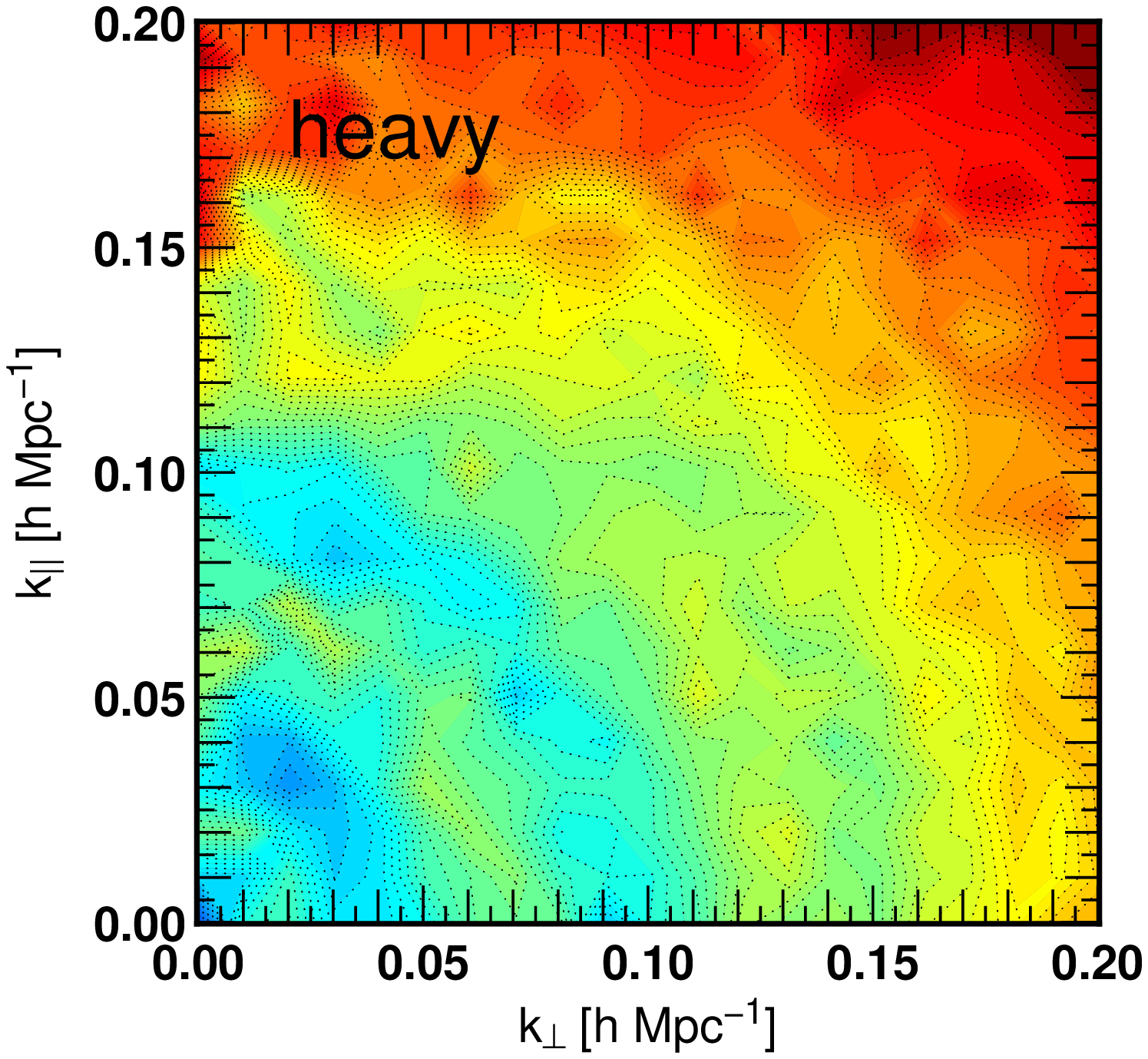} 
   \includegraphics[height=5.4cm]{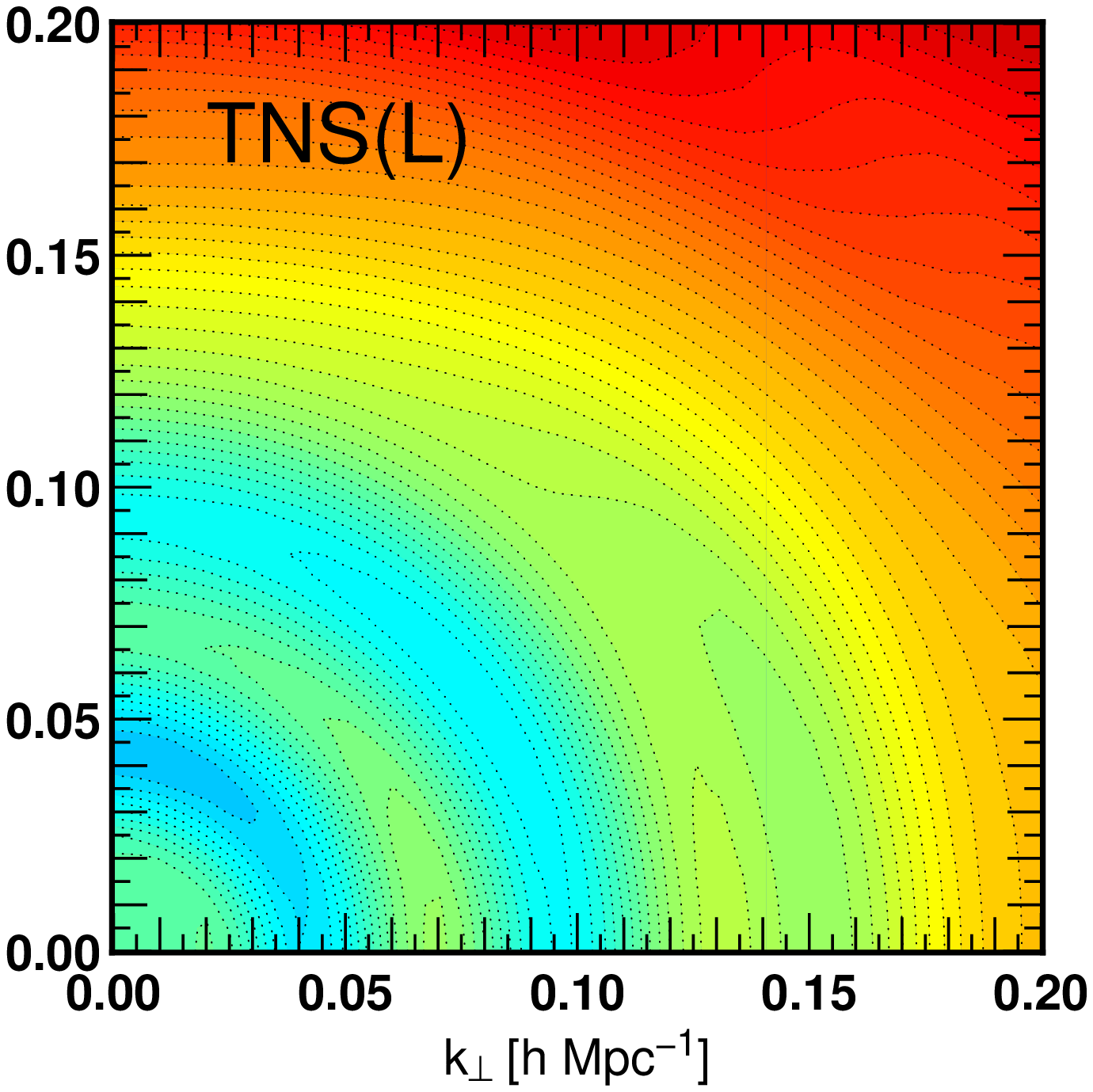} 
   \includegraphics[height=5.4cm]{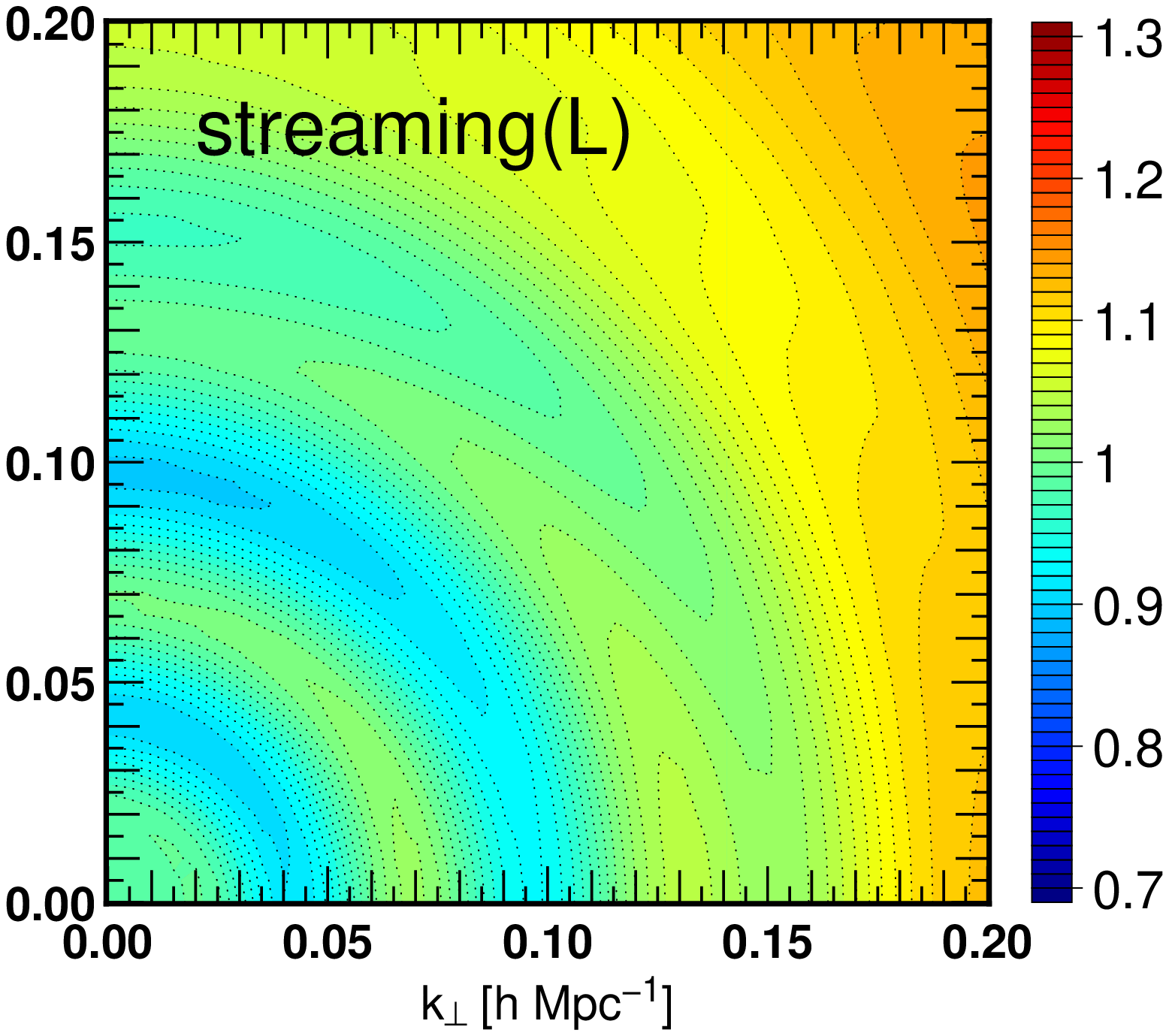} 
   \caption{Power spectra in redshift space normalized by $[b(k)+f\mu^2]^2 \Pnw(k)$ in $(k_\perp,k_\parallel)$ plane.
   We adopt the function $b(k)$ in the normalization factor for halos directly measured in 
   real space by Eq.~(\ref{eq:b_from_pk}), 
   while we set $b(k)=1$ for dark matter.
   Results for dark matter, {\tt light} and {\tt heavy} halo subsamples are shown
   from top to bottom panels: 
   measurements from the N-body simulations (left), 
   best-fit TNS model (middle), 
   best-fit streaming model (right).
   Note that in computing model predictions, we adopt 
   the Lorentzian form of the Finger-of-God factor for the model predictions, and the velocity dispersion $\sigma_{\rm v}$
   is determined by fitting the model predictions to the corresponding N-body data shown in the left panels.}
   \label{fig:2d}
\end{figure*}

Let us first show Fig.~\ref{fig:2d}, in which we plot the ratio of 
the power spectra defined by 
\be
\frac{P(k,\mu)}{\{b(k)+f\,\mu^2\}^2P_{\rm nw}(k)}
\label{eq:ratio_pk_2D}
\ee
for the dark matter (top), {\tt light} (middle), and {\tt heavy} (bottom) 
halos. The results are shown in the $(k_\perp,k_\parallel)$ plane, 
where $k_\perp$ and $k_\parallel$  are the wavenumber 
perpendicular and parallel to the line-of-sight direction, and are related to 
the quantities $k$ and $\mu$ through 
$k_\perp=k(1-\mu^2)^{1/2}$ and $k_\parallel=k\,\mu$. 
Note again that the quantities $P_{\rm nw}$ and $f$ 
mean the smooth reference linear spectrum based on 
the no-wiggle approximation by Ref.~\cite{EH98}, and growth-rate parameter 
defined by $f=d\ln D_+/d\ln a$, respectively.

In each panel of Fig.~\ref{fig:2d}, the results obtained from the 
N-body simulations, the analytic models (\ref{eq:TNS_halo}) with 
and without corrections (indicated by `TNS' and `streaming', respectively) 
are presented from left to right. Here, the analytic model predictions 
are plotted adopting the best-fit values of $\sigma_{\rm v}$, which are 
derived based on the Lorentzian form of the Finger-of-God damping.  
In all of the panels, a ring-like structure 
originated from the  BAOs is clearly manifest, and 
the acoustic feature is rather prominent along the $k_\perp$ axis, 
where redshift distortions have no effect. On the other hand, 
in the line-of-sight direction, the appreciable reduction of the 
amplitude is found 
in the dark matter power spectrum, while no such effect is apparent 
in the halo power spectra. Rather, the {\tt heavy} halo subsample exhibits 
a bit large enhancement at $k_\parallel\gtrsim0.1\,h$Mpc$^{-1}$, 
and the effect becomes significant as increasing the wavenumber. 
Note that with the ratio defined by Eq.~(\ref{eq:ratio_pk_2D}), 
the linear Kaiser effect, which also enhances the 
clustering amplitude, is effectively eliminated in Fig.~\ref{fig:2d}. 
While a damping seen in the dark matter clustering would be due to 
the Finger-of-God effect, the absence of damping or a large enhancement  
in the halo clustering indicates that the Finger-of-God damping seems 
ineffective, and/or the new effect of redshift distortions additionally 
arises and it may compensate for the Finger-of-God damping. Since 
the clustering amplitude is rather sensitive to the halo subsamples, 
this new effect seems to have a strong dependence on the clustering bias.

Now, looking at the analytic model predictions (\ref{eq:TNS_halo}) 
with and without the corrections, there appears a noticeable 
difference between the two models in the result of {\tt heavy} 
halo subsample. Remarkably, the TNS model [i.e., 
Eq.~(\ref{eq:TNS_halo}) including the corrections] reproduces 
a large enhancement seen in the N-body simulations quite well. 
Since the main distinction between the two models comes from 
the presence of additional terms given in Eqs.~(\ref{eq:A_term}) and 
(\ref{eq:B_term}), the enhancement found in the N-body simulations 
can be interpreted as a non-linear effect of redshift distortions,  
characterized by the correction terms. Indeed, 
the corrections originated from the $A$ and $B$ terms 
exhibit a strong dependence on the 
halo bias parameter $b$, and this fact naturally explains the 
sensitive dependence of the clustering enhancement on the halo subsamples. 
These findings are basically consistent with the recent claims by 
Ref.~\cite{Tang11,Reid11} (see also Ref.~\cite{Saito11}),
and the present paper further reveals an impact of the non-linear redshift distortions 
on the power spectrum in two dimensions.

\begin{figure*}[!t] 
   \begin{center}
   \includegraphics[width=15.5cm]{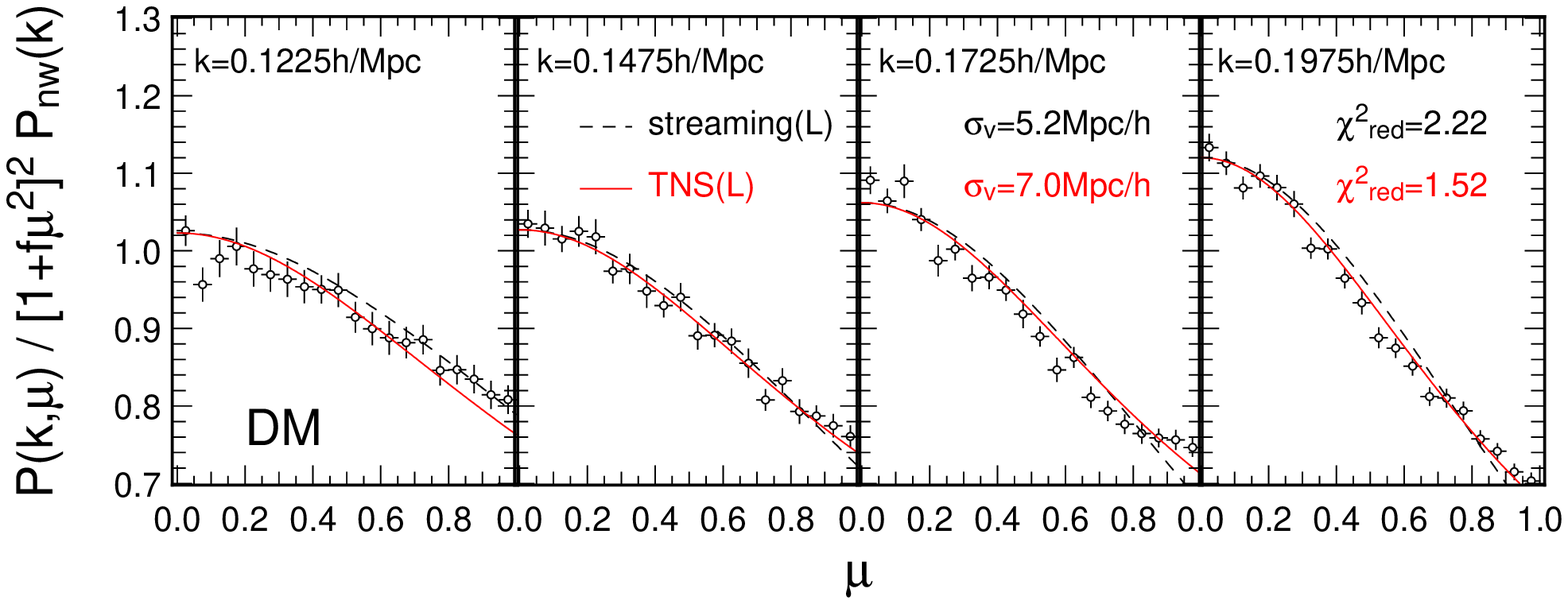} 
   \includegraphics[width=15.5cm]{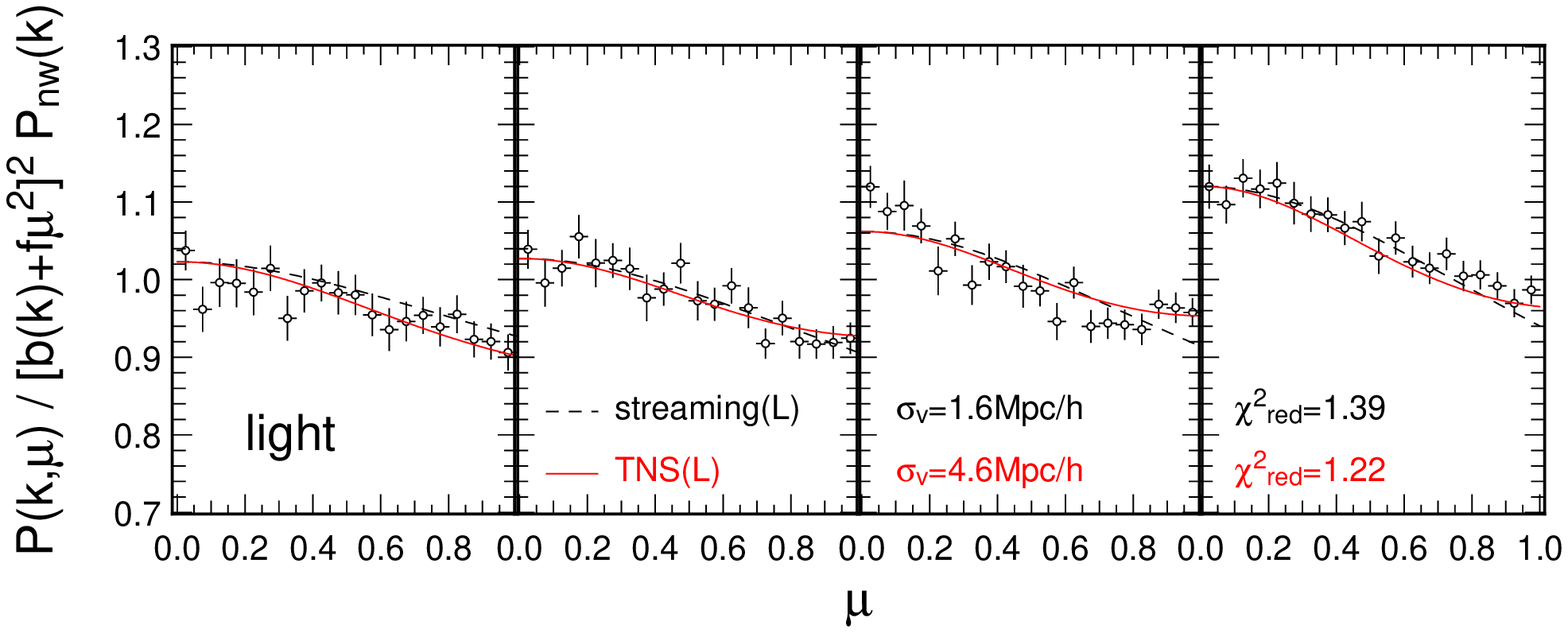}
   \includegraphics[width=15.5cm]{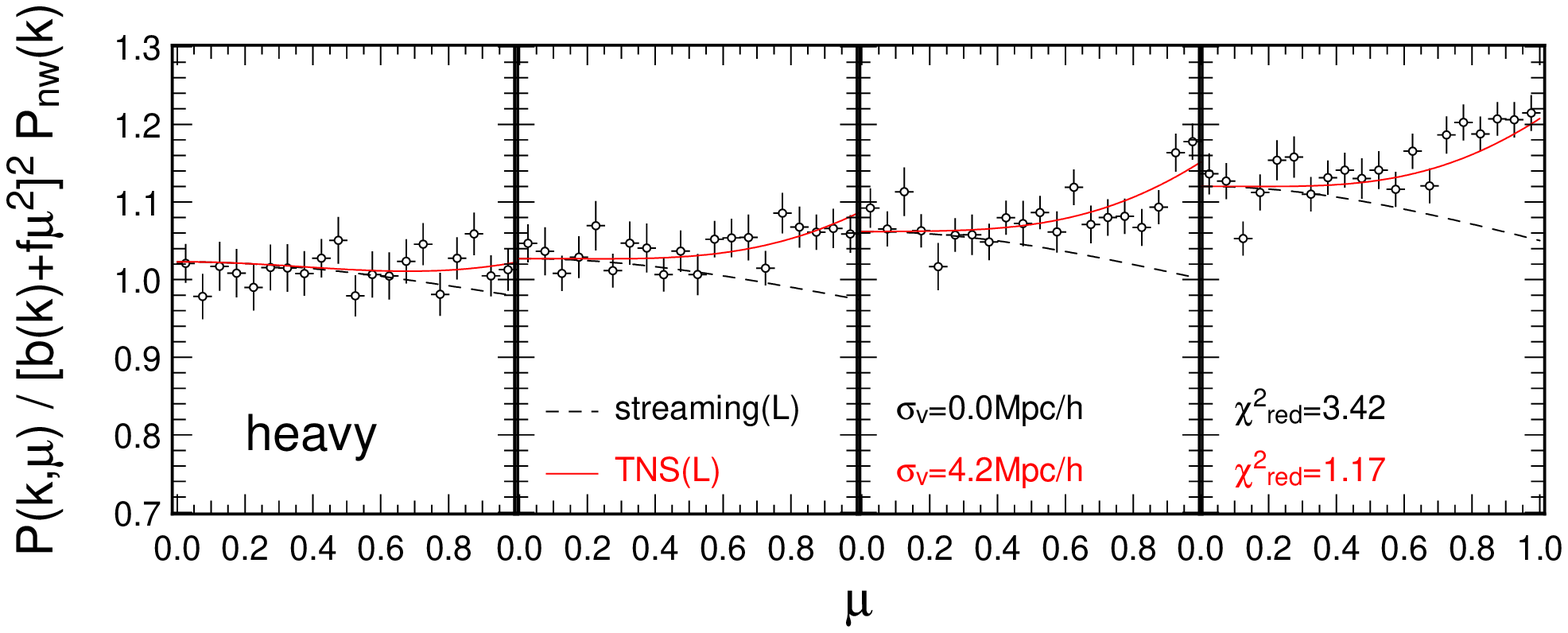} 
   \end{center}
   \caption{Power spectra in redshift space normalized by $[b(k)+f\mu^2]^2 \Pnw(k)$
     as a function of directional cosine $\mu$ for fixed wavenumbers, $k=0.1225, 0.1475, 0.1725$ and
     $0.1975\,h$Mpc$^{-1}$ (from left to right).
     We adopt the function $b(k)$ in the normalization factor for halos directly measured 
     in real space by Eq.~(\ref{eq:b_from_pk}), while we set $b(k)=1$ for dark matter.
     The symbols are measurements from N-body simulations with errorbars estimated 
     by Eq.~(\ref{eq:pk_error}), while the 
     lines are fits by the formula of Eq.~(\ref{eq:TNS_halo}) with (solid, labeled by `TNS(L)') and 
     without (dashed, labeled by `streaming(L)') correction terms. {\it Top}: dark matter. 
     {\it Middle}: {\tt light} halo subsample. {\it Bottom}: {\tt heavy} halo subsample.
     Note that we adopt the Lorentzian form of the Finger-of-God factor for the model predictions, and
     the velocity dispersion $\sigma_{\rm v}$ is determined by fitting to the corresponding
     N-body data. The best-fit values of $\sigma_{\rm v}$ as well as the resultant reduced
     chi-squared defined by Eq.~(\ref{eq:chisquared}) for each model are shown in the panels.}
   \label{fig:k-mu}
\end{figure*}

In Fig.~\ref{fig:k-mu},  
for a more quantitative comparison of the analytic models 
with N-body simulations, we fix the wavenumber $k$, and plot the 
ratio (\ref{eq:ratio_pk_2D}) as a function of directional cosine $\mu$; 
dark matter, {\tt light} and {\tt heavy} halo subsample, from top to bottom. 
In each panel, the N-body data are shown in symbols, 
and the results of best-fit analytic models with and without corrections,  
indicated by 
`TNS' and `streaming', are depicted as the solid and dashed lines. 
For reference, the fitted results of velocity dispersion $\sigma_{\rm v}$ 
and the goodness of fit $\chi_{\rm red}^2$ are also shown.  
Compared to the dark matter clustering, a large enhancement 
seen in the {\tt heavy} subsample is statistically significant,  
and the discrepancy between simulations and the streaming model is 
clearly evident. Even in the absence of Finger-of-God damping (i.e., 
$\sigma_{\rm v}=0$), the streaming model cannot reproduce the
N-body trend. By contrast, the model including the corrections 
(i.e., the TNS model) successfully explains the N-body results. 
Although no appreciable 
difference of the results between the two models is found by eye 
in the dark matter and {\tt light} subsample, the resultant  
goodness of fit for the TNS model is better than that of 
the streaming model, statistically indicating that the model 
including the corrections 
successfully describes a real physical effect of redshift 
distortions seen in the N-body simulations. This point will be 
further investigated in detail in subsequent subsections.

\subsection{Multipole expansion}
\label{subsec:multipole}
\begin{figure*}[!t] 
   \centering
   \includegraphics[height=5.4cm]{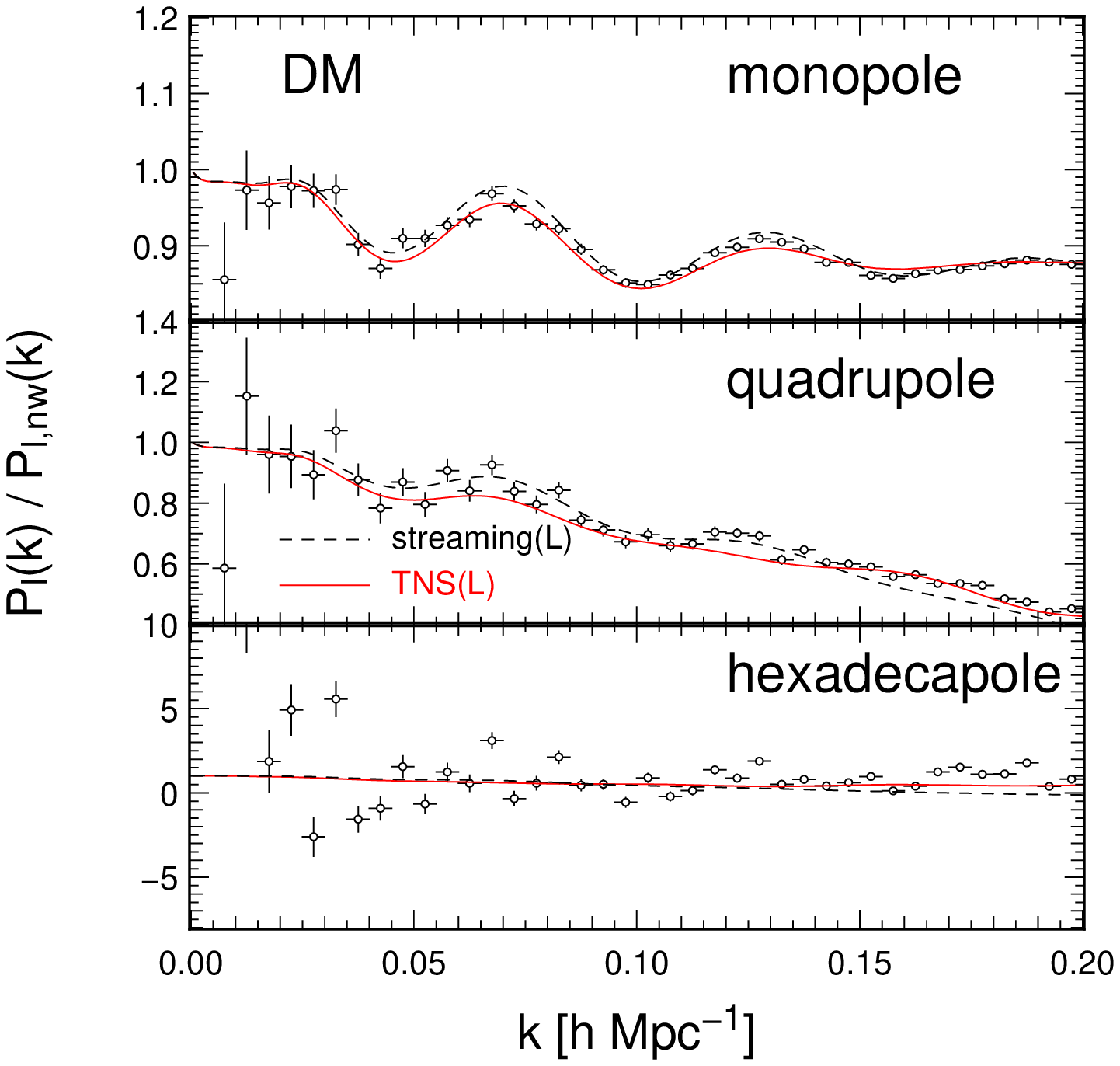} 
   \includegraphics[height=5.4cm]{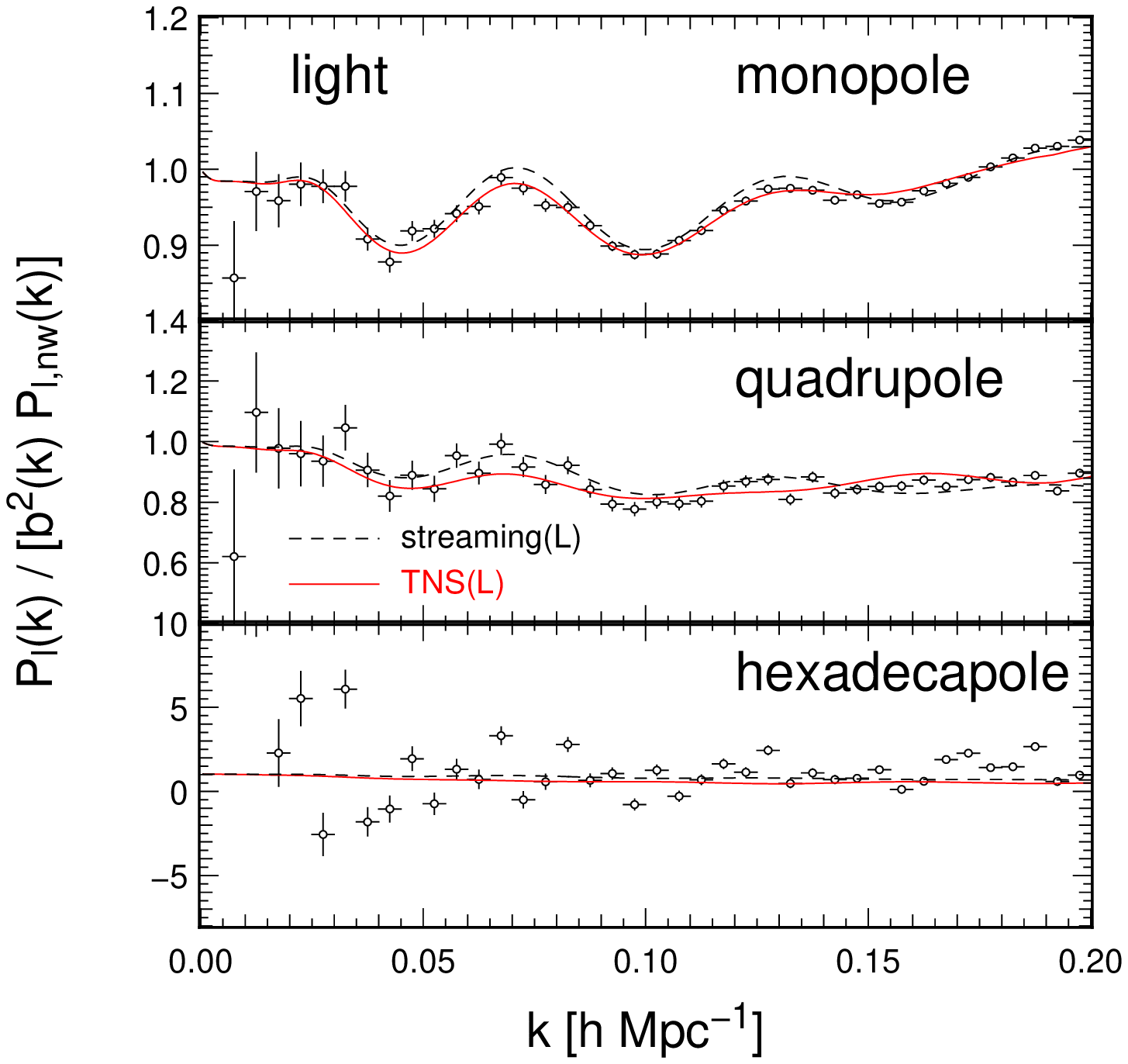} 
   \includegraphics[height=5.4cm]{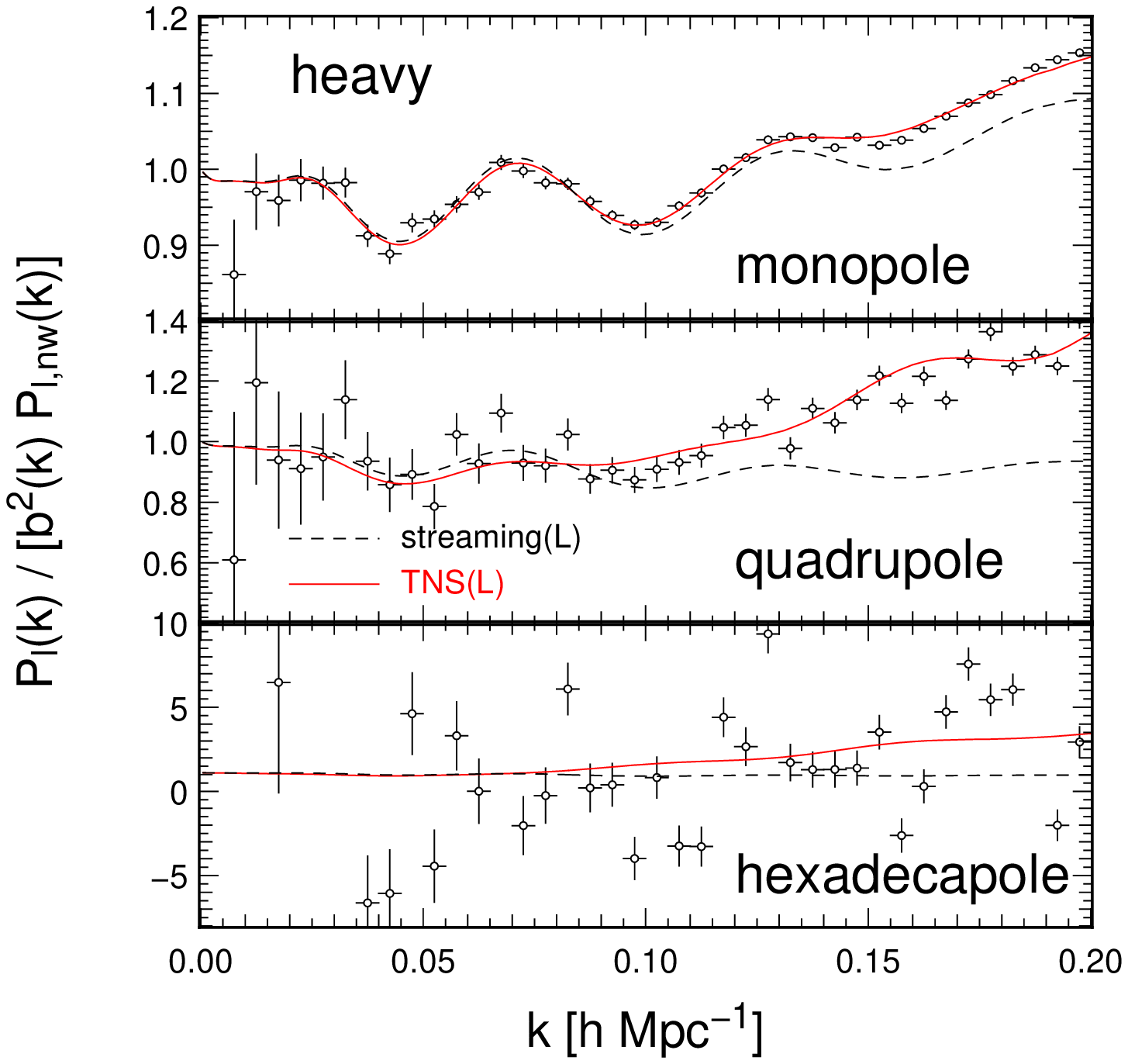} 
   \caption{Multipole power spectra normalized by the smooth reference 
     power spectrum $b^2(k)P_{\ell,{\rm nw}}(k)$ [see Eq.~(\ref{eq:P_ell_nw}) for
     the definition]. The monopole,  
   quadrupole, and hexadecapole are shown from top to bottom. 
   The symbols show the results of the N-body simulations, while
   the lines represent the best-fit TNS (solid) and streaming (dashed) models.
   {\it Left}: dark matter. {\it Middle}: {\tt light} halo subsample. {\it Right}: {\tt heavy} halo subsample.
   Note that we adopt the Lorentzian form of the Finger-of-God factor for the model predictions,
   and the velocity dispersion $\sigma_{\rm v}$ is determined by fitting to the N-body data.}
   \label{fig:multi}
\end{figure*}

The multipole expansion is an alternative technique for quantifying 
the clustering anisotropies, and it has been frequently applied in 
the literature to extract the cosmological information from 
the redshift-space power spectrum \cite{Hamilton:1997zq}. 
Especially, the lower-multipole 
spectra are shown to be powerful to measure the 
characteristic scales of BAOs (e.g., \cite{Yamamoto:2008gr,Padmanabhan:2008ag,Taruya11}). It is also helpful for 
our purpose to understand visually 
how well the analytic models can accurately 
reproduce the simulation results. 
Given the redshift-space power spectrum in two dimensions, $P(k,\mu)$, 
the multipole power spectrum, $P_\ell(k)$, is defined as, 
\be
P(k,\mu) &=& \sum_{\ell = {\rm even}}P_\ell(k)\,
\mathcal{P}_\ell(\mu),\\
P_\ell(k) &=& \frac{2\ell+1}2\int_{-1}^{1}\rmd\mu \,
\mathcal{P}_\ell(\mu)\,P(k,\mu),
\label{eq:multipole}
\ee
where $\mathcal{P}_\ell(\mu)$ denotes the Legendre polynomial.

Fig.~\ref{fig:multi} shows the monopole ($\ell=0$), quadrupole ($\ell=2$), 
and hexadecapole ($\ell=4$) power spectra (from top to bottom). The results
are shown for dark matter, {\tt light}, and {\tt heavy} halo subsamples 
(from left to right), divided by the 
smooth reference power spectrum, $b^2(k)\,P_{\ell,{\rm nw}}(k)$, where 
the power spectrum $P_{\ell,{\rm nw}}(k)$ is computed from $P_{\rm nw}(k)$ 
taking the linear Kaiser effect into account:  
\begin{align}
P_{\ell,{\rm nw}}(k) = \left\{
\begin{array}{ccl}
(1 + \frac{2}{3}\beta + \frac{1}{5}\beta^2) P_{\rm nw}(k) &;& \ell=0 
\\
\\
(\frac{4}{3}\beta + \frac{4}{7}\beta^2) P_{\rm nw}(k) &;& \ell=2 
\\
\\
\frac{8}{35}\beta^2 P_{\rm nw}(k) &;& \ell=4 
\end{array}
\right.
\label{eq:P_ell_nw}
\end{align}
with the quantity $\beta$ defined by $\beta(k)=f/b(k)$.

Apart from the noisy hexadecapole power spectrum, which is 
largely due to the finite sampling of the Fourier modes along the 
$\mu$ direction, both of the streaming and TNS models 
are broadly consistent with the N-body results for 
the dark matter and {\tt light} halo subsample. A closer 
look at the acoustic feature 
reveals a slight discrepancy between the simulations 
and streaming model (dashed). Further, 
for the {\tt heavy} subsample, the streaming model fails to reproduce 
the enhancement of clustering amplitude on small scales. 
These are all what we found in Figs.~\ref{fig:2d} and \ref{fig:k-mu}, 
and are consistent with Ref.~\cite{Taruya10}. 
The notable point is that these discrepancies are visually 
evident even for the dark matter and {\tt light} halos. That is,  
while the streaming model predicts a rather clear 
BAO signal, the actual acoustic structure seen in the N-body 
simulations seems somewhat degraded, and because of this, the streaming 
model slightly overshoots the N-body results at 
low-$k$, and eventually turn to underestimate at high-$k$. 
By contrast, the model including the corrections 
reproduces the N-body results fairly well, and quantitatively explains 
a slightly smeared BAOs in the cases of dark matter and {\tt light} halos,  
as well as a large enhancement in the {\tt heavy} subsample.

Carefully looking at the results of 
the TNS model, however, 
the oscillatory behavior seen in the quadrupole spectrum seems to 
be rather over-smeared, and the acoustic structure 
becomes featureless. Because of that, the 
visual impression for the agreement with N-body simulations is 
somewhat degraded. As it has been discussed in 
Ref.~\cite{Taruya10}, this is presumably due to our
heterogeneous treatment on the corrections $A$ and $B$ using the 
standard PT calculations. The standard PT is known to generically 
give a strong suppression on the acoustic feature in BAOs. 
As gravitational clustering develops, it may incorrectly lead to  
a phase reversal of BAOs 
(e.g., \cite{Jeong06,Nishimichi09}).
Although the corrections $A$ and $B$ 
are basically small, at $z=0.35$ of our samples, the non-linearity of 
gravitational clustering is strong, and the application of 
standard PT might be subtle. Nevertheless, as we will show later, 
the goodness of fit inferred 
from $\chi_{\rm red}^2$ favors the model including the 
corrections, and the model broadly gives a good agreement with 
N-body simulations in all halo subsamples. In this respect, 
the model (\ref{eq:TNS_halo}) captures an important aspect of 
redshift-space clustering, and the role of the 
non-linear coupling described by $A$ and $B$ terms is 
quite essential.

\subsection{Dependence on halo mass and maximum wavenumber}
\label{subsec:accuracy}
\begin{figure*}[!t] 
   \centering
   \includegraphics[height=8.4cm]{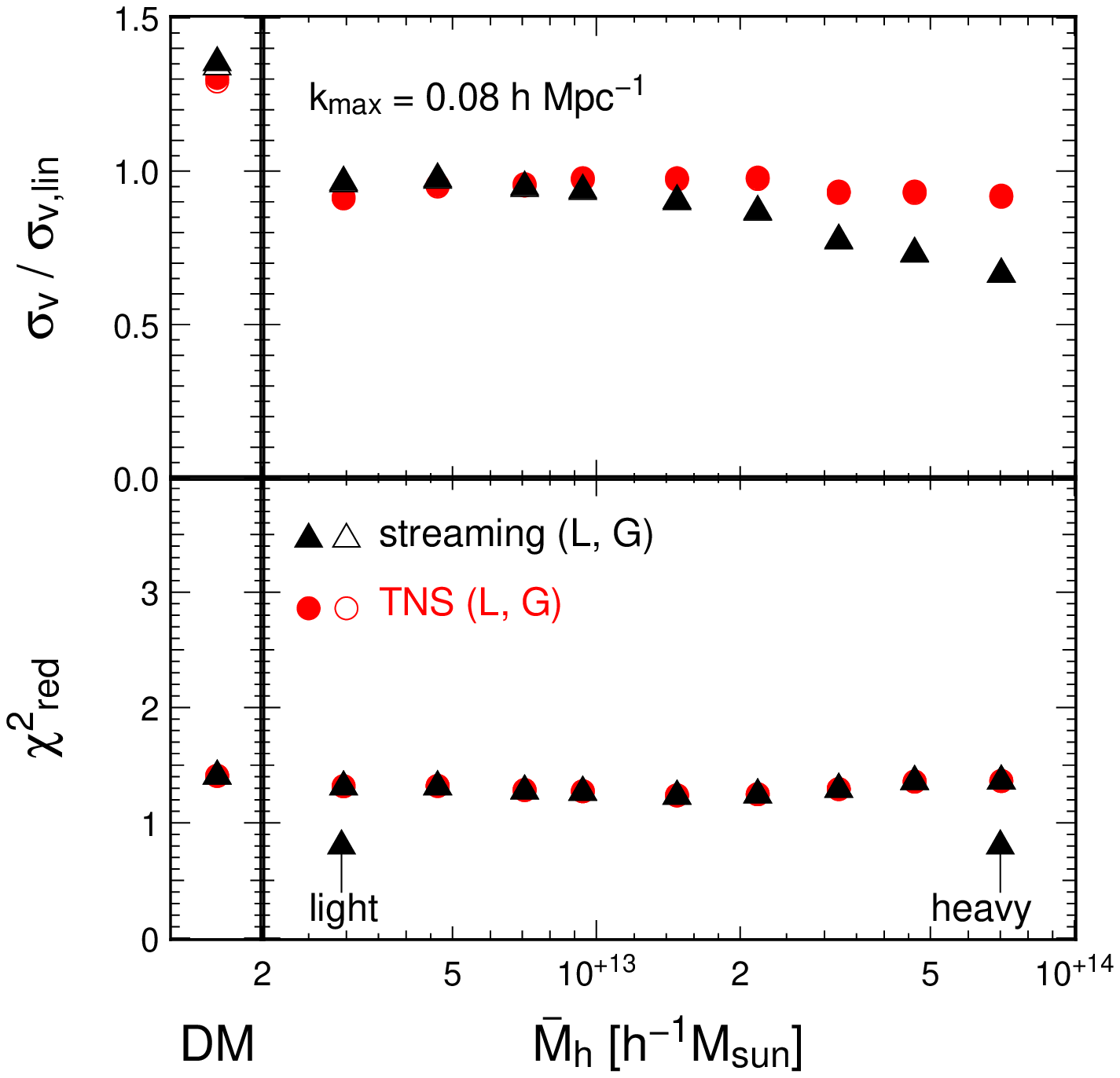} 
   \includegraphics[height=8.4cm]{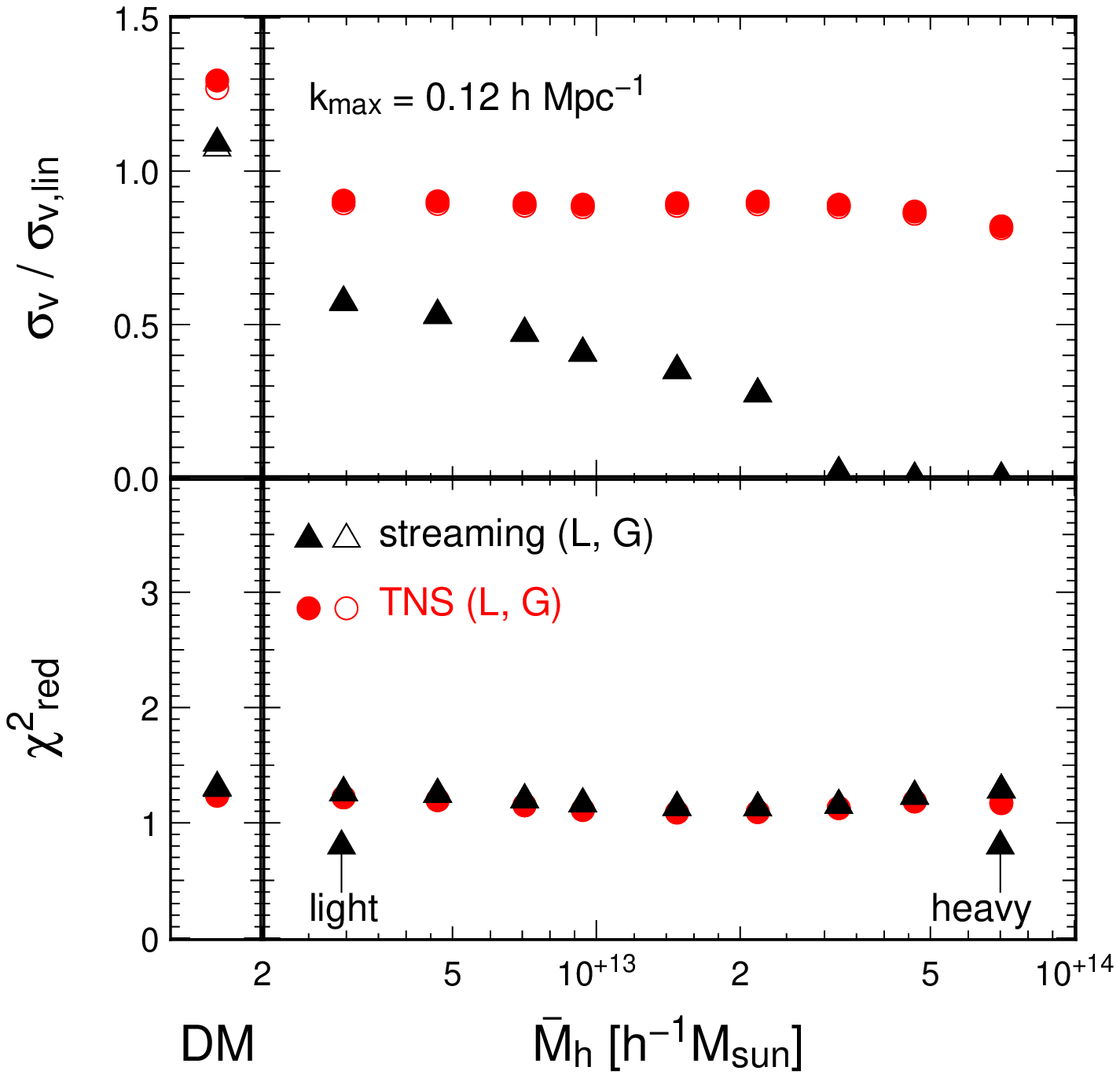} 
   \includegraphics[height=8.4cm]{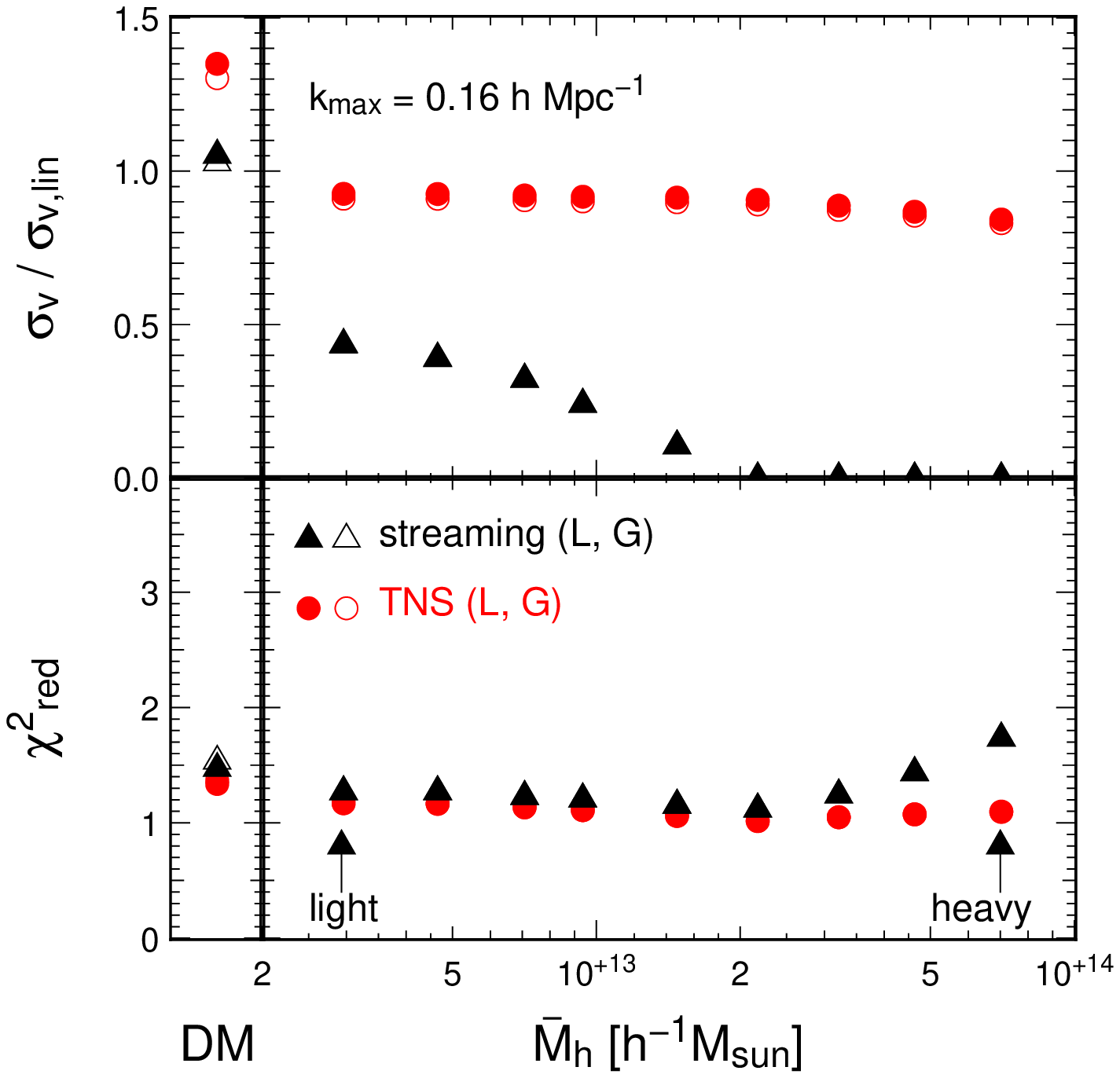} 
   \includegraphics[height=8.4cm]{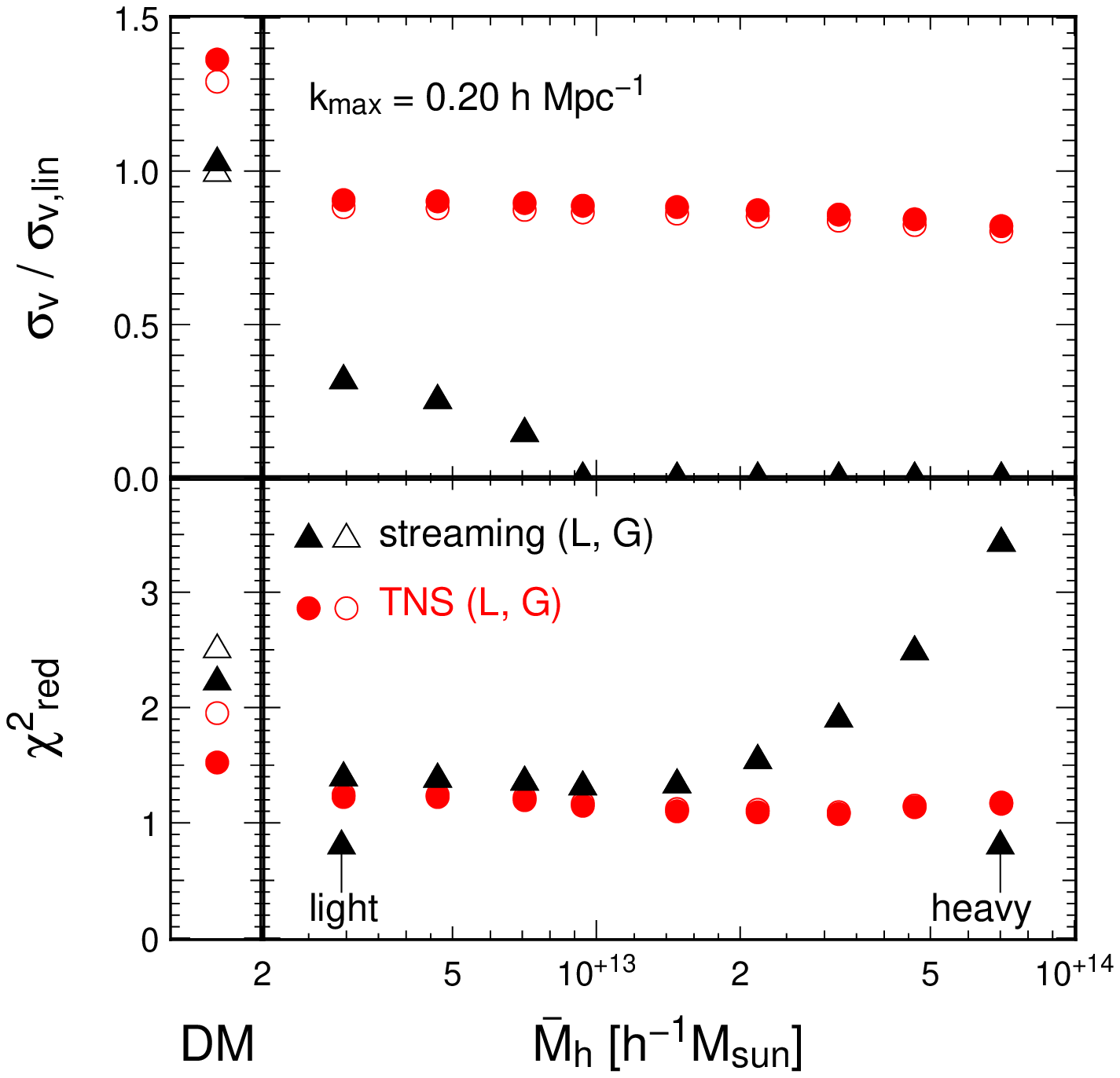} 
   \caption{Best-fit parameter $\sigma_{\rm v}$ divided by the linear
   theory prediction $\sigma_{\rm v,lin}^2\equiv \int dq P_{\rm lin}(q) / (6\pi^2)$
   and the resultant reduced chi-squared (\ref{eq:chisquared})
   for various dark matter/halo samples and maximum wavenumbers. 
   The triangles represent the results from 
   the streaming model, while the circles are from the TNS model. In computing the analytic 
   models, we examine both the Lorentzian and Gaussian forms of the 
   Finger-of-God damping (indicated by filled and open symbols, and labeled by 
   'L' and 'G' in the figure legend, respectively). 
   The maximum wavenumber in the fitting, $k_{\rm max}$, is set to 
   $0.08$, $0.12$, $0.16$ and $0.2h$Mpc$^{-1}$, from top left to bottom right.}
   \label{fig:param_s-space}
\end{figure*}

So far, we have focused on the specific halo subsamples, {\tt light} and 
{\tt heavy}, as well as dark matter, and explored the 
sample dependent properties of the redshift-space power spectrum. 
Here, we examine all the halo subsamples, and 
investigate the halo mass 
dependence on the validity of analytic models in a more quantitative way, 
by evaluating the goodness of fit, $\chi^2_{\rm red}$.

Fig.~\ref{fig:param_s-space} summarizes the comparison between 
analytic models and N-body simulations. In each panel, varying the 
maximum wavenumber $k_{\rm max}$ for the range of the fitting, the resultant 
best-fit value of $\sigma_{\rm v}$ divided by the linear theory prediction 
$\sigma_{\rm v,lin}^2\equiv \int dq P_{\rm lin}(q) / (6\pi^2)$ (top) as well as the reduced 
chi-squared $\chi^2_{\rm red}$ (bottom) 
are shown in each analytic model. Here, in computing the analytic 
models, we examine both the Lorentzian and Gaussian forms of the 
Finger-of-God damping (indicated by filled and open symbols, and labeled by 
'L' and 'G' in the figure legend, respectively). 
The results are all plotted as a function of the averaged halo mass 
$\overline{M}_h$ for each subsample.

We first notice that the best-fit value $\sigma_{\rm v}$ and 
$\chi_{\rm red}^2$ are 
irrelevant for the choice of the Finger-of-God damping function, 
consistent with the result found in Ref.~\cite{Taruya10}.  
This is presumably because the maximum wavenumber $k_{\rm max}$ for 
the range of fitting is limited, $k_{\rm max}\lesssim0.2\,h$Mpc$^{-1}$, 
and no appreciable effect of the small-scale clustering appears 
on the scales of our interest. On the other hand, the best-fit 
$\sigma_{\rm v}$ shows a strong model dependence. While the resultant  
numerical values decrease with the halo mass in the streaming model,   
the fitted results of $\sigma_{\rm v}$ 
mostly remain unchanged in the TNS model, and interestingly  
are very close to the linear theory prediction $\sigma_{\rm v,lin}$.     
Further, the best-fit value in the streaming model is sensitively 
affected by the range of fitting, and for $k_{\rm max}\gtrsim0.1\,h$Mpc$^{-1}$, 
it significantly deviates from the linear theory prediction, and 
eventually becomes zero for massive halo subsamples. 
This peculiar behavior is somewhat counter-intuitive, 
and indicates the breakdown of the model prescription.

Indeed, for the streaming model, 
the goodness of fit indicated by $\chi^2_{\rm red}$ becomes worse 
as increasing the halo mass and maximum wavenumber. Apparently, 
even at the best-fit values $\sigma_{\rm v}=0$, 
the reduced chi-squared $\chi^2_{\rm red}$ tends to keep the similar 
values to those obtained in the TNS model, and 
the streaming model illusively seems to 
reproduce the N-body results as well. Note, however, that the goodness 
of fit indicated by $\chi^2_{\rm red}$ is merely a 
statistical measure for the discrepancy between model and data, 
and cannot be used to judge which model is physically plausible or not. 
As it has been demonstrated in Ref.~\cite{Taruya10} (see also 
Ref.~\cite{Taruya11}), the parameter estimation based on the streaming model 
can cause a large systematic bias in the best-fit values of the growth-rate 
parameter $f$, while the model including the corrections correctly reproduces
the fiducial values fairly well within the statistical 
error at 1-$\sigma$ level. Although this demonstration has been made 
in the dark matter case, the same would be true for 
the halo subsamples. Since the velocity dispersion directly measured from 
N-body simulations is known to be nearly independent of the scales and halo 
mass (e.g., \cite{Jenkins98,Hamana03}), 
we conclude that the streaming model fails to capture some aspects 
of redshift-space clustering, and the model including the corrections 
can provide a better prescription which also gives a physically reasonable 
estimate of the halo velocity statistics.

\section{Implications}
\label{sec:discussion}

The goal of accurately modeling redshift distortions 
is to provide a practically useful theoretical template, and    
with a precision power spectrum measurement, we can tightly 
constrain and/or measure
the growth history of large-scale structure as well as the 
cosmological distances.  
For on-going and/or future galaxy surveys, the precision of the power spectrum 
measurement is expected to reach at a percent-level, and even a slightly improper 
modeling of the redshift-space power spectrum 
can lead to a serious systematic bias in the cosmological parameter 
estimations. Then, an important question is how seriously 
the tiny deficit seen in the 
streaming model can lead to a biased estimation of cosmological parameters.

Here, assuming that the model including the corrections 
$A$ and $B$, given by Eq.~(\ref{eq:TNS_halo}), is correct, 
we estimate the size of systematic biases 
caused by the incorrect model of redshift distortions,  
particularly focusing on the growth-rate parameter. To do this, we 
use the Fisher matrix formalism presented in Ref.~\cite{Taruya10,Taruya11}. 
Provided the survey parameters that characterize the galaxy surveys such as 
the observed redshift $z_c$, survey volume $V$, and number density of 
galaxies $n_g$, the Fisher matrix analysis enables us to estimate not only 
the statistical uncertainty but also 
the systematic bias for the best-fit value of cosmological parameters 
around the fiducial cosmological model. Following the 
Fisher matrix calculations given by Ref.~\cite{Taruya10,Taruya11}, 
we specifically consider the on-going and up-coming galaxy surveys, 
BOSS \cite{Schlegel:2009hj}, SuMIRe-PFS \cite{Suto10}, and 
HETDEX \cite{Hill:2008mv}, and derive the expected constraints on  
the growth-rate parameter. Table \ref{tab:survey}
summarizes the properties of the three galaxy surveys used in the 
Fisher matrix calculations.

For illustrative purpose, we assume a
prior knowledge of the real-space power spectra, $P_{\delta\delta}$, 
$P_{\delta\theta}$, and $P_{\theta\theta}$. 
Then, adopting the streaming model 
as a wrong prior assumption for the template spectrum, we try to estimate 
the growth-rate parameter $f$ from the redshift-space power spectrum in 
two dimensions. Notice that in addition to redshift distortions, 
the imperfect knowledge of the late-time cosmic expansion leads to 
another appreciable effect of clustering anisotropies,  
known as the Alcock-Paczynski effect. 
Here, we marginalize over the Alcock-Paczynski effect characterized by 
the distance scales, $D_A(z)/D_{A,{\rm fid}}$ and $H(z)/H_{\rm fid}$ 
(e.g., \cite{Ballinger:1996cd,Magira:1999bn,Taruya11}), where 
the subscript $_{\rm fid}$ indicates the fiducial value. As a result, 
the total number of free parameters assumed in the parameter estimation
is five, i.e., $f$, $D_A(z)/D_{\rm fid}$, $H(z)/H_{\rm fid}$, $b$, and 
$\sigma_{\rm v}$. For each galaxy sample obtained from the three surveys, 
we assume the scale-independent linear bias 
listed in Table~\ref{tab:survey}, and adopt the linear theory estimate, 
$\sigma_{\rm v,lin}$, as a fiducial value of the velocity 
dispersion, $\sigma_{\rm v}$.

\begin{figure}[t!] 
   \includegraphics[width=7.8cm]{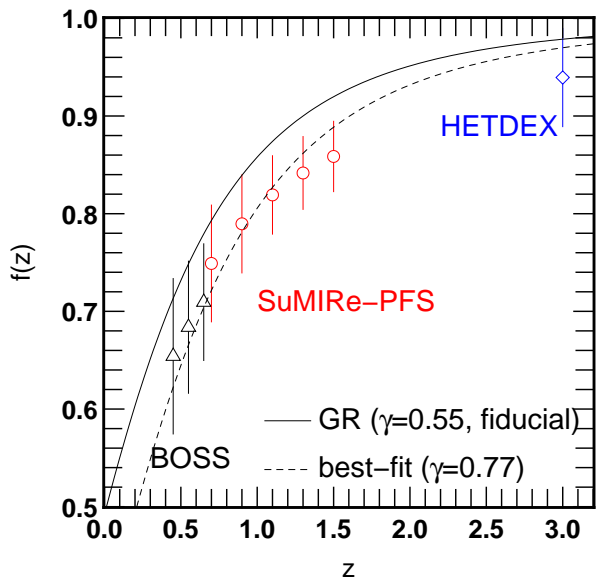} 
   \caption{Expected constraints on the growth-rate parameter $f(z)$ around the best-fit values (symbols), 
   from three on-going/planned surveys.
   The results are obtained based on the Fisher matrix formalism assuming the streaming model
   as a wrong prior of the redshift-space power spectrum.
   The fiducial cosmology is depicted as the solid line, while the dashed line shows the
   best-fit model when we parametrize as Eq.~(\ref{eq:gamma}).}
   \label{fig:forecast}
\end{figure}

Fig.~\ref{fig:forecast} summarizes the forecast result for the Fisher matrix 
analysis based on the power spectrum measurements in two dimensions. 
The fiducial growth-rate parameter is depicted as solid line, and  
the errorbar around each symbol represents the expected 1-$\sigma$ 
constraint around the biased 
best-fit value of $f$, which results from the Fisher matrix calculation
for each redshift subsample. Note that 
the maximum wavenumber $k_{\rm max}$ used in the parameter estimation 
is chosen so as to be well within the applicable range of 
improved PT (see Table~\ref{tab:survey}), 
where the non-linear gravitational clustering is 
still moderate, and can only change the real-space power spectrum 
by $5\sim10\%$.

As it is clearly seen in Fig.~\ref{fig:forecast}, 
the estimated best-fit values of 
$f$ are systematically lower than the fiducial values. 
Apparently, the deviation from the fiducial value 
is not so large at each point, and the biased estimates of the 
best-fit value are sometimes inside the 1-$\sigma$ error around the 
fiducial value. However, the combined result with all the constraints  
strongly disfavors the fiducial model. For instance, if we parametrize 
the growth-rate parameter by (e.g., \cite{Peebles:1980,Linder05})
\be
f(z) = \left[\Omega_{\rm m}(z)\right]^\gamma,
\label{eq:gamma}
\ee
we obtain $\gamma=0.77\pm0.04$, which significantly deviates from 
$\gamma=0.55$ expected from the fiducial model.

The reason why the theoretical template neglecting the 
corrections leads to the underestimation of growth-rate parameter 
is basically explained by the mismatch of the overall shape of 
the power spectra. As shown in Sec.~\ref{subsec:multipole}, 
on large scales, 
the best-fit streaming model slightly overtakes 
the amplitude of each multipole spectrum in both the TNS model 
and N-body simulations, and then the model eventually turns to underestimate 
at high-$k$. Note that these are the outcome of the single-parameter 
fit. In order to closely match the TNS model 
and/or N-body simulations, a simple way is to further adjust 
the parameters other than $\sigma_{\rm v}$. 
Among the remaining four parameters, 
the bias parameter $b$, 
and distance scales $D_A$ and $H$ are indeed well-constrained, and 
their contribution to the systematic bias is likely to be rather small.  
In this respect, 
a slight change of the growth-rate parameter, which controls the 
strength of Kaiser effect, is the only possible 
way to match the result of TNS model and N-body simulations on large scales. 
Note that whether the streaming model overestimates or 
underestimates the parameter $f(z)$ heavily 
depends on what scales one weighs in the parameter estimation. 
In our specific example in Fig.~\ref{fig:forecast}, 
we choose rather conservative values 
of $k_{\rm max}$. Thus, the net effect of the systematic bias appears as an 
underestimation of $f$. A great emphasis is that even the  
tiny discrepancy seen at low-$k$ (see Fig.~\ref{fig:multi}) 
can lead to a serious systematic bias in the on-going and/or up-coming 
power spectrum measurements.

Recently, the WiggleZ dark energy survey has provided a large data set of 
redshift-space clustering around $0.1<z<0.9$ \cite{Blake11a}, 
from which tight constraints on the growth-rate parameter have been 
put based on the streaming model.  
After an extensive test for various models of redshift distortions, 
the authors of Ref.~\cite{Blake11a} conclude that 
both the streaming model combining 
the fitting formula by \cite{Jennings11} and the TNS model 
provide a reasonable fit to the observed clustering data, and can be 
used to derive the constraints on the growth-rate parameter. 
Indeed, the clustering bias of this galaxy sample 
is shown to be rather small $b\sim1$, and no appreciable distinction between  
the TNS and streaming models is manifest within the statistical errors.  
In this sense, a reasonable goodness of fit in both the streaming and TNS models 
sounds rather consistent with our findings, and the derived results on 
the growth-rate parameter would be unbiased. However, 
galaxy samples with a large clustering bias 
$b>1$ will certainly exhibit a non-negligible effect of non-linear 
redshift distortions, and a tiny deficit in the theoretical template 
can lead to a large systematic bias on $f$, 
as shown in Fig.~\ref{fig:forecast}. 
Hence, the suitable choice of the theoretical template is very crucial for 
future precision measurements of the power spectrum.

\begin{table}[b]
\begin{ruledtabular}
\caption{Survey parameters adopted in the Fisher matrix analyses. The parameters, 
$z_c$, $V$, $n_g$, $b$, and $k_{\rm max}$, represent the survey volume, galaxy number density, bias and
the maximum wavenumber included in the analysis, respectively.
The units are in $h^{-3}$Gpc$^3$ for $V$, $h^3$Mpc$^{-3}$ for $n_g$ and $h$Mpc$^{-1}$ for $k_{\rm max}$.}
\label{tab:survey}
\begin{tabular}{c||c|c|c|c|c}
 & $z_c$ & $V$ & $n_g$ & $b$ & $k_{\rm max}$\\
 \hline\hline
BOSS & $0.45$ & $1.1$ & $3\times10^{-4}$ & $2.2$ & $0.15$\\
           & $0.55$ & $1.5$ & $3\times10^{-4}$ & $2.2$ & $0.15$\\
           & $0.65$ & $1.9$ & $3\times10^{-4}$ & $2.2$ & $0.15$\\
\hline
SuMIRe-PFS & $0.7$ & $0.8$ & $3\times10^{-4}$ & $1.5$ & $0.2$\\
           & $0.9$ & $1.1$ & $3\times10^{-4}$ & $1.5$ & $0.2$\\
           & $1.1$ & $1.4$ & $4\times10^{-4}$ & $1.5$ & $0.2$\\
           & $1.3$ & $1.6$ & $4\times10^{-4}$ & $1.5$ & $0.2$\\
           & $1.5$ & $1.7$ & $4\times10^{-4}$ & $1.5$ & $0.2$\\
\hline
HETDEX & $3.0$ & $3.0$ & $2.5\times10^{-4}$ & $2.5$ & $0.4$
\end{tabular}
\end{ruledtabular}
\end{table}

\section{Summary}
\label{sec:summary}

We have presented the redshift-space power spectra of dark matter and 
halos measured from a large set of cosmological N-body simulations, and
tested the analytical models of redshift distortions 
against the N-body results. With a particular 
focus on the redshift distortion effects of halos, 
we have created the nine halo catalogs over a wide mass range.
The resultant volume and signal-to-noise ratio of each catalog are roughly
comparable to those in the SDSS DR7 LRG.

We found that the measured halo power spectrum in redshift space exhibits  
a large enhancement in amplitude on large scales, and the effect becomes 
significant as increasing the halo mass. This enhancement cannot be simply 
explained by the Kaiser effect, and the popular model of redshift 
distortions called streaming model fails to reproduce the N-body results. 
The detailed comparison with N-body simulations further reveals 
that even for the less massive halos with $M_{\rm h}\gtrsim10^{13}h^{-1}M_\odot$, 
a small but non-negligible discrepancy is manifest on large scales,   
and the best-fit values of the velocity dispersion $\sigma_{\rm v}$ is 
rather sensitive to the range of fitting and halo subsamples. 
By contrast, the model based on the perturbation theory description 
(i.e., the TNS model), which includes non-trivial corrections to the streaming 
model, gives a better agreement with N-body simulations for every halo 
subsample. In particular, the model quantitatively 
explains a large enhancement of the power spectrum amplitude in 
{\tt heavy} subsample fairy well. These results indicates that the 
non-linear coupling between density and velocity fields induces 
the new effect of redshift distortions, which seems to have 
sensitive dependence on the clustering bias. The corrections in the 
TNS model (\ref{eq:TNS_halo}) can describe the observed feature of 
the redshift-space halo clustering very well.

To investigate how the slightly improper modeling of redshift distortions 
affects the cosmological parameter estimation, we have also estimated 
the size of systematic bias based on the Fisher matrix formalism.  
Especially focusing on the growth rate parameter, 
we consider the on-going and upcoming power spectrum measurements 
from BOSS, SuMIRe-PFS and HETDEX, and found that 
even the small deficit in the theoretical template of the power spectrum 
can produce a large systematic bias in the growth rate parameter, and 
the fiducial model would be erroneously ruled out. 
This is true even if we choose a conservative value of $k_{\rm max}$. 
Hence, the accurate description of the redshift-space power spectrum is 
quite essential, and the model including the corrections can be used 
as a reliable theoretical template of the power spectrum.

Finally, we leave the following issue as a future work. Throughout the 
analysis, we have calibrated the halo bias relationship directly 
from the N-body simulations in real space, and the measured result of 
the bias parameter has been used to compute the analytic model of 
the redshift-space power spectrum, after subtracting the non-Poissonian 
residual noise. Of course, this treatment is infeasible in real 
observations, and we must combine other data set in order to get the 
information of the clustering bias. One plausible approach is to 
combine the weak lensing measurement. Since the weak lensing measurement   
directly probes the matter distribution, the cross correlation between 
weak lensing and galaxy clustering data can give a powerful way 
to simultaneously characterize the clustering bias and the residual noise 
contributions. Synergy with the weak lensing measurement is worth considering,  
and the methodology to extract a pure clustering signal should be 
further exploited.

\acknowledgments
This work was initiated while staying in Institut d'Astrophysique de Paris 
and Institut de Physique Th\'eorique. We thank Thierry Sousbie, Sebastien 
Peirani, Stephane Colombi, Francis Bernardeau and Patrick Valageas for 
their hospitality and discussions during our stay.
We also thank Masahiro Takada, Issha Kayo and Eiichiro Komatsu 
for their fruitful comments and discussions. 
T.~N. is supported by a Grant-in-Aid for Japan Society for 
the Promotion of Science (JSPS) Fellows (PD: 22-181) and by 
World Premier International Research
Center Initiative (WPI Initiative), MEXT, Japan. 
A.~T. is supported 
by a Grant-in-Aid for Scientific Research from JSPS (No. 21740168).
Numerical calculations for the present work have been in part carried 
out under the ''Interdisciplinary Computational Science Program'' in 
Center for Computational Sciences, University 
of Tsukuba, and also on Cray XT4 at Center for Computational
Astrophysics, CfCA, of National Astronomical Observatory of Japan.  
This work was supported in part by 
Grant-in-Aid for Scientific Research on Priority Areas No.~467 
``Probing the Dark Energy through an Extremely Wide and Deep Survey with 
Subaru Telescope'', and JSPS Core-to-Core Program ``International 
Research Network for Dark Energy''.



\begin{thebibliography}{65}
\expandafter\ifx\csname natexlab\endcsname\relax\def\natexlab#1{#1}\fi
\expandafter\ifx\csname bibnamefont\endcsname\relax
  \def\bibnamefont#1{#1}\fi
\expandafter\ifx\csname bibfnamefont\endcsname\relax
  \def\bibfnamefont#1{#1}\fi
\expandafter\ifx\csname citenamefont\endcsname\relax
  \def\citenamefont#1{#1}\fi
\expandafter\ifx\csname url\endcsname\relax
  \def\url#1{\texttt{#1}}\fi
\expandafter\ifx\csname urlprefix\endcsname\relax\def\urlprefix{URL }\fi
\providecommand{\bibinfo}[2]{#2}
\providecommand{\eprint}[2][]{\url{#2}}

\bibitem[{\citenamefont{Hamilton}(Kluwer Academic, Dordrecht, The Netherlands,
  1998)}]{Hamilton:1997zq}
\bibinfo{author}{\bibfnamefont{A.~J.~S.} \bibnamefont{Hamilton}},
  \bibinfo{journal}{{in {\it The Evolving Universe}, edited by D.~Hamilton}}
  pp. \bibinfo{pages}{pp. 185--275} (\bibinfo{year}{Kluwer Academic, Dordrecht,
  The Netherlands, 1998}), \eprint{astro-ph/9708102}.

\bibitem[{\citenamefont{Peebles}(Princeton University Press,
  1980)}]{Peebles:1980}
\bibinfo{author}{\bibfnamefont{P.}~\bibnamefont{Peebles}},
  \bibinfo{journal}{{\it The large-scale structure of the universe}}
  (\bibinfo{year}{Princeton University Press, 1980}).

\bibitem[{\citenamefont{Linder}(2008)}]{Linder:2007nu}
\bibinfo{author}{\bibfnamefont{E.~V.} \bibnamefont{Linder}},
  \bibinfo{journal}{Astropart. Phys.} \textbf{\bibinfo{volume}{29}},
  \bibinfo{pages}{336} (\bibinfo{year}{2008}), \eprint{0709.1113}.

\bibitem[{\citenamefont{Guzzo et~al.}(2008)}]{Guzzo:2008ac}
\bibinfo{author}{\bibfnamefont{L.}~\bibnamefont{Guzzo}} \bibnamefont{et~al.},
  \bibinfo{journal}{Nature} \textbf{\bibinfo{volume}{451}},
  \bibinfo{pages}{541} (\bibinfo{year}{2008}), \eprint{0802.1944}.

\bibitem[{\citenamefont{Yamamoto et~al.}(2008)\citenamefont{Yamamoto, Sato, and
  Huetsi}}]{Yamamoto:2008gr}
\bibinfo{author}{\bibfnamefont{K.}~\bibnamefont{Yamamoto}},
  \bibinfo{author}{\bibfnamefont{T.}~\bibnamefont{Sato}}, \bibnamefont{and}
  \bibinfo{author}{\bibfnamefont{G.}~\bibnamefont{Huetsi}},
  \bibinfo{journal}{Prog. Theor. Phys.} \textbf{\bibinfo{volume}{120}},
  \bibinfo{pages}{609} (\bibinfo{year}{2008}), \eprint{0805.4789}.

\bibitem[{\citenamefont{Song and Percival}(2009)}]{Song:2008qt}
\bibinfo{author}{\bibfnamefont{Y.-S.} \bibnamefont{Song}} \bibnamefont{and}
  \bibinfo{author}{\bibfnamefont{W.~J.} \bibnamefont{Percival}},
  \bibinfo{journal}{JCAP} \textbf{\bibinfo{volume}{0910}}, \bibinfo{pages}{004}
  (\bibinfo{year}{2009}), \eprint{0807.0810}.

\bibitem[{\citenamefont{Song and Kayo}(2010)}]{Song:2010bk}
\bibinfo{author}{\bibfnamefont{Y.-S.} \bibnamefont{Song}} \bibnamefont{and}
  \bibinfo{author}{\bibfnamefont{I.}~\bibnamefont{Kayo}}
  (\bibinfo{year}{2010}), \eprint{1003.2420}.

\bibitem[{\citenamefont{Kaiser}(1987)}]{Kaiser:1987qv}
\bibinfo{author}{\bibfnamefont{N.}~\bibnamefont{Kaiser}},
  \bibinfo{journal}{Mon. Not. Roy. Astron. Soc.}
  \textbf{\bibinfo{volume}{227}}, \bibinfo{pages}{1} (\bibinfo{year}{1987}).

\bibitem[{\citenamefont{{Hamilton}}(1992)}]{1992ApJ385L5H}
\bibinfo{author}{\bibfnamefont{A.~J.~S.} \bibnamefont{{Hamilton}}},
  \bibinfo{journal}{Astrophys. J.} \textbf{\bibinfo{volume}{385}},
  \bibinfo{pages}{L5} (\bibinfo{year}{1992}).

\bibitem[{\citenamefont{{Alcock} and {Paczynski}}(1979)}]{Alcock79}
\bibinfo{author}{\bibfnamefont{C.}~\bibnamefont{{Alcock}}} \bibnamefont{and}
  \bibinfo{author}{\bibfnamefont{B.}~\bibnamefont{{Paczynski}}},
  \bibinfo{journal}{Nature} \textbf{\bibinfo{volume}{281}},
  \bibinfo{pages}{358} (\bibinfo{year}{1979}).

\bibitem[{\citenamefont{Seo and Eisenstein}(2003)}]{Seo:2003pu}
\bibinfo{author}{\bibfnamefont{H.-J.} \bibnamefont{Seo}} \bibnamefont{and}
  \bibinfo{author}{\bibfnamefont{D.~J.} \bibnamefont{Eisenstein}},
  \bibinfo{journal}{Astrophys. J.} \textbf{\bibinfo{volume}{598}},
  \bibinfo{pages}{720} (\bibinfo{year}{2003}), \eprint{astro-ph/0307460}.

\bibitem[{\citenamefont{Blake and Glazebrook}(2003)}]{Blake:2003rh}
\bibinfo{author}{\bibfnamefont{C.}~\bibnamefont{Blake}} \bibnamefont{and}
  \bibinfo{author}{\bibfnamefont{K.}~\bibnamefont{Glazebrook}},
  \bibinfo{journal}{Astrophys. J.} \textbf{\bibinfo{volume}{594}},
  \bibinfo{pages}{665} (\bibinfo{year}{2003}), \eprint{astro-ph/0301632}.

\bibitem[{\citenamefont{Glazebrook and Blake}(2005)}]{Glazebrook:2005mb}
\bibinfo{author}{\bibfnamefont{K.}~\bibnamefont{Glazebrook}} \bibnamefont{and}
  \bibinfo{author}{\bibfnamefont{C.}~\bibnamefont{Blake}},
  \bibinfo{journal}{Astrophys. J.} \textbf{\bibinfo{volume}{631}},
  \bibinfo{pages}{1} (\bibinfo{year}{2005}), \eprint{astro-ph/0505608}.

\bibitem[{\citenamefont{Shoji et~al.}(2009)\citenamefont{Shoji, Jeong, and
  Komatsu}}]{Shoji:2008xn}
\bibinfo{author}{\bibfnamefont{M.}~\bibnamefont{Shoji}},
  \bibinfo{author}{\bibfnamefont{D.}~\bibnamefont{Jeong}}, \bibnamefont{and}
  \bibinfo{author}{\bibfnamefont{E.}~\bibnamefont{Komatsu}},
  \bibinfo{journal}{Astrophys. J.} \textbf{\bibinfo{volume}{693}},
  \bibinfo{pages}{1404} (\bibinfo{year}{2009}), \eprint{0805.4238}.

\bibitem[{\citenamefont{Padmanabhan and White}(2008)}]{Padmanabhan:2008ag}
\bibinfo{author}{\bibfnamefont{N.}~\bibnamefont{Padmanabhan}} \bibnamefont{and}
  \bibinfo{author}{\bibfnamefont{M.~J.} \bibnamefont{White}, \bibfnamefont{1}},
  \bibinfo{journal}{Phys. Rev.} \textbf{\bibinfo{volume}{D77}},
  \bibinfo{pages}{123540} (\bibinfo{year}{2008}), \eprint{0804.0799}.

\bibitem[{\citenamefont{{Eisenstein} et~al.}(2005)\citenamefont{{Eisenstein},
  {Zehavi}, {Hogg}, {Scoccimarro}, {Blanton}, {Nichol}, {Scranton}, {Seo},
  {Tegmark}, {Zheng} et~al.}}]{Eisenstein05}
\bibinfo{author}{\bibfnamefont{D.~J.} \bibnamefont{{Eisenstein}}},
  \bibinfo{author}{\bibfnamefont{I.}~\bibnamefont{{Zehavi}}},
  \bibinfo{author}{\bibfnamefont{D.~W.} \bibnamefont{{Hogg}}},
  \bibinfo{author}{\bibfnamefont{R.}~\bibnamefont{{Scoccimarro}}},
  \bibinfo{author}{\bibfnamefont{M.~R.} \bibnamefont{{Blanton}}},
  \bibinfo{author}{\bibfnamefont{R.~C.} \bibnamefont{{Nichol}}},
  \bibinfo{author}{\bibfnamefont{R.}~\bibnamefont{{Scranton}}},
  \bibinfo{author}{\bibfnamefont{H.-J.} \bibnamefont{{Seo}}},
  \bibinfo{author}{\bibfnamefont{M.}~\bibnamefont{{Tegmark}}},
  \bibinfo{author}{\bibfnamefont{Z.}~\bibnamefont{{Zheng}}},
  \bibnamefont{et~al.}, \bibinfo{journal}{\apj} \textbf{\bibinfo{volume}{633}},
  \bibinfo{pages}{560} (\bibinfo{year}{2005}), \eprint{arXiv:astro-ph/0501171}.

\bibitem[{\citenamefont{{Cole} et~al.}(2005)\citenamefont{{Cole}, {Percival},
  {Peacock}, {Norberg}, {Baugh}, {Frenk}, {Baldry}, {Bland-Hawthorn},
  {Bridges}, {Cannon} et~al.}}]{Cole05}
\bibinfo{author}{\bibfnamefont{S.}~\bibnamefont{{Cole}}},
  \bibinfo{author}{\bibfnamefont{W.~J.} \bibnamefont{{Percival}}},
  \bibinfo{author}{\bibfnamefont{J.~A.} \bibnamefont{{Peacock}}},
  \bibinfo{author}{\bibfnamefont{P.}~\bibnamefont{{Norberg}}},
  \bibinfo{author}{\bibfnamefont{C.~M.} \bibnamefont{{Baugh}}},
  \bibinfo{author}{\bibfnamefont{C.~S.} \bibnamefont{{Frenk}}},
  \bibinfo{author}{\bibfnamefont{I.}~\bibnamefont{{Baldry}}},
  \bibinfo{author}{\bibfnamefont{J.}~\bibnamefont{{Bland-Hawthorn}}},
  \bibinfo{author}{\bibfnamefont{T.}~\bibnamefont{{Bridges}}},
  \bibinfo{author}{\bibfnamefont{R.}~\bibnamefont{{Cannon}}},
  \bibnamefont{et~al.}, \bibinfo{journal}{Mon. Not. Roy. Astron. Soc.}
  \textbf{\bibinfo{volume}{362}}, \bibinfo{pages}{505} (\bibinfo{year}{2005}),
  \eprint{arXiv:astro-ph/0501174}.

\bibitem[{\citenamefont{{H{\"u}tsi}}(2006)}]{Huetsi06}
\bibinfo{author}{\bibfnamefont{G.}~\bibnamefont{{H{\"u}tsi}}},
  \bibinfo{journal}{A\&A} \textbf{\bibinfo{volume}{449}}, \bibinfo{pages}{891}
  (\bibinfo{year}{2006}), \eprint{arXiv:astro-ph/0512201}.

\bibitem[{\citenamefont{{Percival} et~al.}(2007)\citenamefont{{Percival},
  {Cole}, {Eisenstein}, {Nichol}, {Peacock}, {Pope}, and
  {Szalay}}}]{Percival07}
\bibinfo{author}{\bibfnamefont{W.~J.} \bibnamefont{{Percival}}},
  \bibinfo{author}{\bibfnamefont{S.}~\bibnamefont{{Cole}}},
  \bibinfo{author}{\bibfnamefont{D.~J.} \bibnamefont{{Eisenstein}}},
  \bibinfo{author}{\bibfnamefont{R.~C.} \bibnamefont{{Nichol}}},
  \bibinfo{author}{\bibfnamefont{J.~A.} \bibnamefont{{Peacock}}},
  \bibinfo{author}{\bibfnamefont{A.~C.} \bibnamefont{{Pope}}},
  \bibnamefont{and} \bibinfo{author}{\bibfnamefont{A.~S.}
  \bibnamefont{{Szalay}}}, \bibinfo{journal}{Mon. Not. Roy. Astron. Soc.}
  \textbf{\bibinfo{volume}{381}}, \bibinfo{pages}{1053} (\bibinfo{year}{2007}),
  \eprint{0705.3323}.

\bibitem[{\citenamefont{{Okumura} et~al.}(2008)\citenamefont{{Okumura},
  {Matsubara}, {Eisenstein}, {Kayo}, {Hikage}, {Szalay}, and
  {Schneider}}}]{Okumura08}
\bibinfo{author}{\bibfnamefont{T.}~\bibnamefont{{Okumura}}},
  \bibinfo{author}{\bibfnamefont{T.}~\bibnamefont{{Matsubara}}},
  \bibinfo{author}{\bibfnamefont{D.~J.} \bibnamefont{{Eisenstein}}},
  \bibinfo{author}{\bibfnamefont{I.}~\bibnamefont{{Kayo}}},
  \bibinfo{author}{\bibfnamefont{C.}~\bibnamefont{{Hikage}}},
  \bibinfo{author}{\bibfnamefont{A.~S.} \bibnamefont{{Szalay}}},
  \bibnamefont{and} \bibinfo{author}{\bibfnamefont{D.~P.}
  \bibnamefont{{Schneider}}}, \bibinfo{journal}{\apj}
  \textbf{\bibinfo{volume}{676}}, \bibinfo{pages}{889} (\bibinfo{year}{2008}),
  \eprint{0711.3640}.

\bibitem[{\citenamefont{{Percival} et~al.}(2010)\citenamefont{{Percival},
  {Reid}, {Eisenstein}, {Bahcall}, {Budavari}, {Frieman}, {Fukugita}, {Gunn},
  {Ivezi{\'c}}, {Knapp} et~al.}}]{Percival10}
\bibinfo{author}{\bibfnamefont{W.~J.} \bibnamefont{{Percival}}},
  \bibinfo{author}{\bibfnamefont{B.~A.} \bibnamefont{{Reid}}},
  \bibinfo{author}{\bibfnamefont{D.~J.} \bibnamefont{{Eisenstein}}},
  \bibinfo{author}{\bibfnamefont{N.~A.} \bibnamefont{{Bahcall}}},
  \bibinfo{author}{\bibfnamefont{T.}~\bibnamefont{{Budavari}}},
  \bibinfo{author}{\bibfnamefont{J.~A.} \bibnamefont{{Frieman}}},
  \bibinfo{author}{\bibfnamefont{M.}~\bibnamefont{{Fukugita}}},
  \bibinfo{author}{\bibfnamefont{J.~E.} \bibnamefont{{Gunn}}},
  \bibinfo{author}{\bibfnamefont{{\v Z}.}~\bibnamefont{{Ivezi{\'c}}}},
  \bibinfo{author}{\bibfnamefont{G.~R.} \bibnamefont{{Knapp}}},
  \bibnamefont{et~al.}, \bibinfo{journal}{Mon. Not. Roy. Astron. Soc.}
  \textbf{\bibinfo{volume}{401}}, \bibinfo{pages}{2148} (\bibinfo{year}{2010}),
  \eprint{0907.1660}.

\bibitem[{\citenamefont{{H{\"u}tsi}}(2010)}]{Huetsi10}
\bibinfo{author}{\bibfnamefont{G.}~\bibnamefont{{H{\"u}tsi}}},
  \bibinfo{journal}{Mon. Not. Roy. Astron. Soc.}
  \textbf{\bibinfo{volume}{401}}, \bibinfo{pages}{2477} (\bibinfo{year}{2010}),
  \eprint{0910.0492}.

\bibitem[{\citenamefont{{Blake}
  et~al.}(2011{\natexlab{a}})\citenamefont{{Blake}, {Davis}, {Poole},
  {Parkinson}, {Brough}, {Colless}, {Contreras}, {Couch}, {Croom}, {Drinkwater}
  et~al.}}]{Blake11b}
\bibinfo{author}{\bibfnamefont{C.}~\bibnamefont{{Blake}}},
  \bibinfo{author}{\bibfnamefont{T.}~\bibnamefont{{Davis}}},
  \bibinfo{author}{\bibfnamefont{G.}~\bibnamefont{{Poole}}},
  \bibinfo{author}{\bibfnamefont{D.}~\bibnamefont{{Parkinson}}},
  \bibinfo{author}{\bibfnamefont{S.}~\bibnamefont{{Brough}}},
  \bibinfo{author}{\bibfnamefont{M.}~\bibnamefont{{Colless}}},
  \bibinfo{author}{\bibfnamefont{C.}~\bibnamefont{{Contreras}}},
  \bibinfo{author}{\bibfnamefont{W.}~\bibnamefont{{Couch}}},
  \bibinfo{author}{\bibfnamefont{S.}~\bibnamefont{{Croom}}},
  \bibinfo{author}{\bibfnamefont{M.~J.} \bibnamefont{{Drinkwater}}},
  \bibnamefont{et~al.}, \bibinfo{journal}{ArXiv e-prints}
  (\bibinfo{year}{2011}{\natexlab{a}}), \eprint{1105.2862}.

\bibitem[{\citenamefont{{Beutler} et~al.}(2011)\citenamefont{{Beutler},
  {Blake}, {Colless}, {Jones}, {Staveley-Smith}, {Campbell}, {Parker},
  {Saunders}, and {Watson}}}]{Beutler11}
\bibinfo{author}{\bibfnamefont{F.}~\bibnamefont{{Beutler}}},
  \bibinfo{author}{\bibfnamefont{C.}~\bibnamefont{{Blake}}},
  \bibinfo{author}{\bibfnamefont{M.}~\bibnamefont{{Colless}}},
  \bibinfo{author}{\bibfnamefont{D.~H.} \bibnamefont{{Jones}}},
  \bibinfo{author}{\bibfnamefont{L.}~\bibnamefont{{Staveley-Smith}}},
  \bibinfo{author}{\bibfnamefont{L.}~\bibnamefont{{Campbell}}},
  \bibinfo{author}{\bibfnamefont{Q.}~\bibnamefont{{Parker}}},
  \bibinfo{author}{\bibfnamefont{W.}~\bibnamefont{{Saunders}}},
  \bibnamefont{and} \bibinfo{author}{\bibfnamefont{F.}~\bibnamefont{{Watson}}},
  \bibinfo{journal}{ArXiv e-prints}  (\bibinfo{year}{2011}),
  \eprint{1106.3366}.

\bibitem[{\citenamefont{{Taruya} et~al.}(2011)\citenamefont{{Taruya}, {Saito},
  and {Nishimichi}}}]{Taruya11}
\bibinfo{author}{\bibfnamefont{A.}~\bibnamefont{{Taruya}}},
  \bibinfo{author}{\bibfnamefont{S.}~\bibnamefont{{Saito}}}, \bibnamefont{and}
  \bibinfo{author}{\bibfnamefont{T.}~\bibnamefont{{Nishimichi}}},
  \bibinfo{journal}{ArXiv e-prints}  (\bibinfo{year}{2011}),
  \eprint{1101.4723}.

\bibitem[{\citenamefont{Schlegel et~al.}(2009)\citenamefont{Schlegel, White,
  and Eisenstein}}]{Schlegel:2009hj}
\bibinfo{author}{\bibfnamefont{D.}~\bibnamefont{Schlegel}},
  \bibinfo{author}{\bibfnamefont{M.}~\bibnamefont{White}}, \bibnamefont{and}
  \bibinfo{author}{\bibfnamefont{D.}~\bibnamefont{Eisenstein}}
  (\bibinfo{collaboration}{with input from the SDSS-III})
  (\bibinfo{year}{2009}), \eprint{0902.4680}.

\bibitem[{\citenamefont{Hill et~al.}(2008)}]{Hill:2008mv}
\bibinfo{author}{\bibfnamefont{G.~J.} \bibnamefont{Hill}} \bibnamefont{et~al.}
  (\bibinfo{year}{2008}), \eprint{0806.0183}.

\bibitem[{\citenamefont{{Suto}}(2010)}]{Suto10}
\bibinfo{author}{\bibfnamefont{Y.}~\bibnamefont{{Suto}}}, in
  \emph{\bibinfo{booktitle}{Society of Photo-Optical Instrumentation Engineers
  (SPIE) Conference Series}} (\bibinfo{year}{2010}), vol.
  \bibinfo{volume}{7733} of \emph{\bibinfo{series}{Presented at the Society of
  Photo-Optical Instrumentation Engineers (SPIE) Conference}},
  \eprint{1007.1256}.

\bibitem[{\citenamefont{Beaulieu et~al.}(2010)}]{Beaulieu:2010qi}
\bibinfo{author}{\bibfnamefont{J.~P.} \bibnamefont{Beaulieu}}
  \bibnamefont{et~al.} (\bibinfo{year}{2010}), \eprint{1001.3349}.

\bibitem[{\citenamefont{Gehrels}(2010)}]{Gehrels:2010fn}
\bibinfo{author}{\bibfnamefont{N.}~\bibnamefont{Gehrels}}
  (\bibinfo{year}{2010}), \eprint{1008.4936}.

\bibitem[{\citenamefont{{Davis} and {Peebles}}(1983)}]{1983ApJ...267..465D}
\bibinfo{author}{\bibfnamefont{M.}~\bibnamefont{{Davis}}} \bibnamefont{and}
  \bibinfo{author}{\bibfnamefont{P.~J.~E.} \bibnamefont{{Peebles}}},
  \bibinfo{journal}{Astrophys. J.} \textbf{\bibinfo{volume}{267}},
  \bibinfo{pages}{465} (\bibinfo{year}{1983}).

\bibitem[{\citenamefont{{Okumura} and {Jing}}(2011)}]{Okumura11}
\bibinfo{author}{\bibfnamefont{T.}~\bibnamefont{{Okumura}}} \bibnamefont{and}
  \bibinfo{author}{\bibfnamefont{Y.~P.} \bibnamefont{{Jing}}},
  \bibinfo{journal}{\apj} \textbf{\bibinfo{volume}{726}}, \bibinfo{pages}{5}
  (\bibinfo{year}{2011}), \eprint{1004.3548}.

\bibitem[{\citenamefont{{Taruya} et~al.}(2010)\citenamefont{{Taruya},
  {Nishimichi}, and {Saito}}}]{Taruya10}
\bibinfo{author}{\bibfnamefont{A.}~\bibnamefont{{Taruya}}},
  \bibinfo{author}{\bibfnamefont{T.}~\bibnamefont{{Nishimichi}}},
  \bibnamefont{and} \bibinfo{author}{\bibfnamefont{S.}~\bibnamefont{{Saito}}},
  \bibinfo{journal}{Physical Review D} \textbf{\bibinfo{volume}{82}},
  \bibinfo{pages}{063522} (\bibinfo{year}{2010}), \eprint{1006.0699}.

\bibitem[{\citenamefont{Hatton and Cole}(1998)}]{Hatton:1997xs}
\bibinfo{author}{\bibfnamefont{S.~J.} \bibnamefont{Hatton}} \bibnamefont{and}
  \bibinfo{author}{\bibfnamefont{S.}~\bibnamefont{Cole}},
  \bibinfo{journal}{Mon. Not. Roy. Astron. Soc.}
  \textbf{\bibinfo{volume}{296}}, \bibinfo{pages}{10} (\bibinfo{year}{1998}),
  \eprint{astro-ph/9707186}.

\bibitem[{\citenamefont{{Taruya} and {Hiramatsu}}(2008)}]{Taruya08}
\bibinfo{author}{\bibfnamefont{A.}~\bibnamefont{{Taruya}}} \bibnamefont{and}
  \bibinfo{author}{\bibfnamefont{T.}~\bibnamefont{{Hiramatsu}}},
  \bibinfo{journal}{Astrophys. J.} \textbf{\bibinfo{volume}{674}},
  \bibinfo{pages}{617} (\bibinfo{year}{2008}), \eprint{0708.1367}.

\bibitem[{\citenamefont{{Taruya} et~al.}(2009)\citenamefont{{Taruya},
  {Nishimichi}, {Saito}, and {Hiramatsu}}}]{Taruya09}
\bibinfo{author}{\bibfnamefont{A.}~\bibnamefont{{Taruya}}},
  \bibinfo{author}{\bibfnamefont{T.}~\bibnamefont{{Nishimichi}}},
  \bibinfo{author}{\bibfnamefont{S.}~\bibnamefont{{Saito}}}, \bibnamefont{and}
  \bibinfo{author}{\bibfnamefont{T.}~\bibnamefont{{Hiramatsu}}},
  \bibinfo{journal}{Physical Review D} \textbf{\bibinfo{volume}{80}},
  \bibinfo{pages}{123503} (\bibinfo{year}{2009}), \eprint{0906.0507}.

\bibitem[{\citenamefont{{Scoccimarro}}(2004)}]{Scoccimarro04}
\bibinfo{author}{\bibfnamefont{R.}~\bibnamefont{{Scoccimarro}}},
  \bibinfo{journal}{Physical Review D} \textbf{\bibinfo{volume}{70}},
  \bibinfo{pages}{083007} (\bibinfo{year}{2004}), \eprint{astro-ph/0407214}.

\bibitem[{\citenamefont{Percival and White}(2009)}]{Percival:2008sh}
\bibinfo{author}{\bibfnamefont{W.~J.} \bibnamefont{Percival}} \bibnamefont{and}
  \bibinfo{author}{\bibfnamefont{M.}~\bibnamefont{White}},
  \bibinfo{journal}{Mon. Not. Roy. Astron. Soc.}
  \textbf{\bibinfo{volume}{393}}, \bibinfo{pages}{297} (\bibinfo{year}{2009}),
  \eprint{0808.0003}.

\bibitem[{\citenamefont{Peacock and Dodds}(1994)}]{Peacock:1993xg}
\bibinfo{author}{\bibfnamefont{J.~A.} \bibnamefont{Peacock}} \bibnamefont{and}
  \bibinfo{author}{\bibfnamefont{S.~J.} \bibnamefont{Dodds}},
  \bibinfo{journal}{Mon. Not. Roy. Astron. Soc.}
  \textbf{\bibinfo{volume}{267}}, \bibinfo{pages}{1020} (\bibinfo{year}{1994}),
  \eprint{astro-ph/9311057}.

\bibitem[{\citenamefont{Park et~al.}(1994)\citenamefont{Park, Vogeley, Geller,
  and Huchra}}]{Park:1994fa}
\bibinfo{author}{\bibfnamefont{C.}~\bibnamefont{Park}},
  \bibinfo{author}{\bibfnamefont{M.~S.} \bibnamefont{Vogeley}},
  \bibinfo{author}{\bibfnamefont{M.~J.} \bibnamefont{Geller}},
  \bibnamefont{and} \bibinfo{author}{\bibfnamefont{J.~P.}
  \bibnamefont{Huchra}}, \bibinfo{journal}{Astrophys. J.}
  \textbf{\bibinfo{volume}{431}}, \bibinfo{pages}{569} (\bibinfo{year}{1994}).

\bibitem[{\citenamefont{Ballinger et~al.}(1996)\citenamefont{Ballinger,
  Peacock, and Heavens}}]{Ballinger:1996cd}
\bibinfo{author}{\bibfnamefont{W.~E.} \bibnamefont{Ballinger}},
  \bibinfo{author}{\bibfnamefont{J.~A.} \bibnamefont{Peacock}},
  \bibnamefont{and} \bibinfo{author}{\bibfnamefont{A.~F.}
  \bibnamefont{Heavens}}, \bibinfo{journal}{Mon. Not. Roy. Astron. Soc.}
  \textbf{\bibinfo{volume}{282}}, \bibinfo{pages}{877} (\bibinfo{year}{1996}),
  \eprint{astro-ph/9605017}.

\bibitem[{\citenamefont{Magira et~al.}(2000)\citenamefont{Magira, Jing, and
  Suto}}]{Magira:1999bn}
\bibinfo{author}{\bibfnamefont{H.}~\bibnamefont{Magira}},
  \bibinfo{author}{\bibfnamefont{Y.~P.} \bibnamefont{Jing}}, \bibnamefont{and}
  \bibinfo{author}{\bibfnamefont{Y.}~\bibnamefont{Suto}},
  \bibinfo{journal}{Astrophys. J.} \textbf{\bibinfo{volume}{528}},
  \bibinfo{pages}{30} (\bibinfo{year}{2000}), \eprint{astro-ph/9907438}.

\bibitem[{\citenamefont{{Springel}}(2005)}]{Springel05}
\bibinfo{author}{\bibfnamefont{V.}~\bibnamefont{{Springel}}},
  \bibinfo{journal}{Mon. Not. Roy. Astron. Soc.}
  \textbf{\bibinfo{volume}{364}}, \bibinfo{pages}{1105} (\bibinfo{year}{2005}),
  \eprint{astro-ph/0505010}.

\bibitem[{\citenamefont{{Scoccimarro}}(1998)}]{Scoccimarro98}
\bibinfo{author}{\bibfnamefont{R.}~\bibnamefont{{Scoccimarro}}},
  \bibinfo{journal}{Mon. Not. Roy. Astron. Soc.}
  \textbf{\bibinfo{volume}{299}}, \bibinfo{pages}{1097} (\bibinfo{year}{1998}),
  \eprint{arXiv:astro-ph/9711187}.

\bibitem[{\citenamefont{{Crocce} et~al.}(2006)\citenamefont{{Crocce},
  {Pueblas}, and {Scoccimarro}}}]{Crocce06}
\bibinfo{author}{\bibfnamefont{M.}~\bibnamefont{{Crocce}}},
  \bibinfo{author}{\bibfnamefont{S.}~\bibnamefont{{Pueblas}}},
  \bibnamefont{and}
  \bibinfo{author}{\bibfnamefont{R.}~\bibnamefont{{Scoccimarro}}},
  \bibinfo{journal}{Mon. Not. Roy. Astron. Soc.}
  \textbf{\bibinfo{volume}{373}}, \bibinfo{pages}{369} (\bibinfo{year}{2006}),
  \eprint{astro-ph/0606505}.

\bibitem[{\citenamefont{{Lewis} et~al.}(2000)\citenamefont{{Lewis},
  {Challinor}, and {Lasenby}}}]{Lewis00}
\bibinfo{author}{\bibfnamefont{A.}~\bibnamefont{{Lewis}}},
  \bibinfo{author}{\bibfnamefont{A.}~\bibnamefont{{Challinor}}},
  \bibnamefont{and}
  \bibinfo{author}{\bibfnamefont{A.}~\bibnamefont{{Lasenby}}},
  \bibinfo{journal}{Astrophys. J.} \textbf{\bibinfo{volume}{538}},
  \bibinfo{pages}{473} (\bibinfo{year}{2000}), \eprint{astro-ph/9911177}.

\bibitem[{\citenamefont{{Komatsu} et~al.}(2009)\citenamefont{{Komatsu},
  {Dunkley}, {Nolta}, {Bennett}, {Gold}, {Hinshaw}, {Jarosik}, {Larson},
  {Limon}, {Page} et~al.}}]{Komatsu09}
\bibinfo{author}{\bibfnamefont{E.}~\bibnamefont{{Komatsu}}},
  \bibinfo{author}{\bibfnamefont{J.}~\bibnamefont{{Dunkley}}},
  \bibinfo{author}{\bibfnamefont{M.~R.} \bibnamefont{{Nolta}}},
  \bibinfo{author}{\bibfnamefont{C.~L.} \bibnamefont{{Bennett}}},
  \bibinfo{author}{\bibfnamefont{B.}~\bibnamefont{{Gold}}},
  \bibinfo{author}{\bibfnamefont{G.}~\bibnamefont{{Hinshaw}}},
  \bibinfo{author}{\bibfnamefont{N.}~\bibnamefont{{Jarosik}}},
  \bibinfo{author}{\bibfnamefont{D.}~\bibnamefont{{Larson}}},
  \bibinfo{author}{\bibfnamefont{M.}~\bibnamefont{{Limon}}},
  \bibinfo{author}{\bibfnamefont{L.}~\bibnamefont{{Page}}},
  \bibnamefont{et~al.}, \bibinfo{journal}{Astrophys. J. Suppl.}
  \textbf{\bibinfo{volume}{180}}, \bibinfo{pages}{330} (\bibinfo{year}{2009}),
  \eprint{0803.0547}.

\bibitem[{\citenamefont{{Eisenstein} et~al.}(2001)\citenamefont{{Eisenstein},
  {Annis}, {Gunn}, {Szalay}, {Connolly}, {Nichol}, {Bahcall}, {Bernardi},
  {Burles}, {Castander} et~al.}}]{Eisenstein01}
\bibinfo{author}{\bibfnamefont{D.~J.} \bibnamefont{{Eisenstein}}},
  \bibinfo{author}{\bibfnamefont{J.}~\bibnamefont{{Annis}}},
  \bibinfo{author}{\bibfnamefont{J.~E.} \bibnamefont{{Gunn}}},
  \bibinfo{author}{\bibfnamefont{A.~S.} \bibnamefont{{Szalay}}},
  \bibinfo{author}{\bibfnamefont{A.~J.} \bibnamefont{{Connolly}}},
  \bibinfo{author}{\bibfnamefont{R.~C.} \bibnamefont{{Nichol}}},
  \bibinfo{author}{\bibfnamefont{N.~A.} \bibnamefont{{Bahcall}}},
  \bibinfo{author}{\bibfnamefont{M.}~\bibnamefont{{Bernardi}}},
  \bibinfo{author}{\bibfnamefont{S.}~\bibnamefont{{Burles}}},
  \bibinfo{author}{\bibfnamefont{F.~J.} \bibnamefont{{Castander}}},
  \bibnamefont{et~al.}, \bibinfo{journal}{Astron. J.}
  \textbf{\bibinfo{volume}{122}}, \bibinfo{pages}{2267} (\bibinfo{year}{2001}),
  \eprint{arXiv:astro-ph/0108153}.

\bibitem[{\citenamefont{{Abazajian} et~al.}(2009)\citenamefont{{Abazajian},
  {Adelman-McCarthy}, {Ag{\"u}eros}, {Allam}, {Allende Prieto}, {An},
  {Anderson}, {Anderson}, {Annis}, {Bahcall} et~al.}}]{Abazajian09}
\bibinfo{author}{\bibfnamefont{K.~N.} \bibnamefont{{Abazajian}}},
  \bibinfo{author}{\bibfnamefont{J.~K.} \bibnamefont{{Adelman-McCarthy}}},
  \bibinfo{author}{\bibfnamefont{M.~A.} \bibnamefont{{Ag{\"u}eros}}},
  \bibinfo{author}{\bibfnamefont{S.~S.} \bibnamefont{{Allam}}},
  \bibinfo{author}{\bibfnamefont{C.}~\bibnamefont{{Allende Prieto}}},
  \bibinfo{author}{\bibfnamefont{D.}~\bibnamefont{{An}}},
  \bibinfo{author}{\bibfnamefont{K.~S.~J.} \bibnamefont{{Anderson}}},
  \bibinfo{author}{\bibfnamefont{S.~F.} \bibnamefont{{Anderson}}},
  \bibinfo{author}{\bibfnamefont{J.}~\bibnamefont{{Annis}}},
  \bibinfo{author}{\bibfnamefont{N.~A.} \bibnamefont{{Bahcall}}},
  \bibnamefont{et~al.}, \bibinfo{journal}{Astrophys. J. Suppl.}
  \textbf{\bibinfo{volume}{182}}, \bibinfo{pages}{543} (\bibinfo{year}{2009}),
  \eprint{0812.0649}.

\bibitem[{\citenamefont{{Hockney} and {Eastwood}}(1981)}]{Hockney81}
\bibinfo{author}{\bibfnamefont{R.~W.} \bibnamefont{{Hockney}}}
  \bibnamefont{and} \bibinfo{author}{\bibfnamefont{J.~W.}
  \bibnamefont{{Eastwood}}}, \emph{\bibinfo{title}{{Computer Simulation Using
  Particles}}} (\bibinfo{year}{1981}).

\bibitem[{\citenamefont{{Seljak} et~al.}(2009)\citenamefont{{Seljak}, {Hamaus},
  and {Desjacques}}}]{Seljak09}
\bibinfo{author}{\bibfnamefont{U.}~\bibnamefont{{Seljak}}},
  \bibinfo{author}{\bibfnamefont{N.}~\bibnamefont{{Hamaus}}}, \bibnamefont{and}
  \bibinfo{author}{\bibfnamefont{V.}~\bibnamefont{{Desjacques}}},
  \bibinfo{journal}{Physical Review Letters} \textbf{\bibinfo{volume}{103}},
  \bibinfo{pages}{091303} (\bibinfo{year}{2009}), \eprint{0904.2963}.

\bibitem[{\citenamefont{{Feldman} et~al.}(1994)\citenamefont{{Feldman},
  {Kaiser}, and {Peacock}}}]{Feldman94}
\bibinfo{author}{\bibfnamefont{H.~A.} \bibnamefont{{Feldman}}},
  \bibinfo{author}{\bibfnamefont{N.}~\bibnamefont{{Kaiser}}}, \bibnamefont{and}
  \bibinfo{author}{\bibfnamefont{J.~A.} \bibnamefont{{Peacock}}},
  \bibinfo{journal}{Astrophys. J.} \textbf{\bibinfo{volume}{426}},
  \bibinfo{pages}{23} (\bibinfo{year}{1994}), \eprint{astro-ph/9304022}.

\bibitem[{\citenamefont{{Eisenstein} and {Hu}}(1998)}]{EH98}
\bibinfo{author}{\bibfnamefont{D.~J.} \bibnamefont{{Eisenstein}}}
  \bibnamefont{and} \bibinfo{author}{\bibfnamefont{W.}~\bibnamefont{{Hu}}},
  \bibinfo{journal}{Astrophys. J.} \textbf{\bibinfo{volume}{496}},
  \bibinfo{pages}{605} (\bibinfo{year}{1998}), \eprint{arXiv:astro-ph/9709112}.

\bibitem[{\citenamefont{{Cooray} and {Sheth}}(2002)}]{Cooray02}
\bibinfo{author}{\bibfnamefont{A.}~\bibnamefont{{Cooray}}} \bibnamefont{and}
  \bibinfo{author}{\bibfnamefont{R.}~\bibnamefont{{Sheth}}},
  \bibinfo{journal}{Phys. Rep.} \textbf{\bibinfo{volume}{372}},
  \bibinfo{pages}{1} (\bibinfo{year}{2002}), \eprint{arXiv:astro-ph/0206508}.

\bibitem[{\citenamefont{{Hamaus} et~al.}(2010)\citenamefont{{Hamaus}, {Seljak},
  {Desjacques}, {Smith}, and {Baldauf}}}]{Hamaus10}
\bibinfo{author}{\bibfnamefont{N.}~\bibnamefont{{Hamaus}}},
  \bibinfo{author}{\bibfnamefont{U.}~\bibnamefont{{Seljak}}},
  \bibinfo{author}{\bibfnamefont{V.}~\bibnamefont{{Desjacques}}},
  \bibinfo{author}{\bibfnamefont{R.~E.} \bibnamefont{{Smith}}},
  \bibnamefont{and}
  \bibinfo{author}{\bibfnamefont{T.}~\bibnamefont{{Baldauf}}},
  \bibinfo{journal}{Phys. Rev. D} \textbf{\bibinfo{volume}{82}},
  \bibinfo{pages}{043515} (\bibinfo{year}{2010}), \eprint{1004.5377}.

\bibitem[{\citenamefont{{Tang} et~al.}(2011)\citenamefont{{Tang}, {Kayo}, and
  {Takada}}}]{Tang11}
\bibinfo{author}{\bibfnamefont{J.}~\bibnamefont{{Tang}}},
  \bibinfo{author}{\bibfnamefont{I.}~\bibnamefont{{Kayo}}}, \bibnamefont{and}
  \bibinfo{author}{\bibfnamefont{M.}~\bibnamefont{{Takada}}},
  \bibinfo{journal}{ArXiv e-prints}  (\bibinfo{year}{2011}),
  \eprint{1103.3614}.

\bibitem[{\citenamefont{{Reid} and {White}}(2011)}]{Reid11}
\bibinfo{author}{\bibfnamefont{B.~A.} \bibnamefont{{Reid}}} \bibnamefont{and}
  \bibinfo{author}{\bibfnamefont{M.}~\bibnamefont{{White}}},
  \bibinfo{journal}{ArXiv e-prints}  (\bibinfo{year}{2011}),
  \eprint{1105.4165}.

\bibitem[{\citenamefont{{Saito} et~al.}(2011)\citenamefont{{Saito}, {Takada},
  and {Taruya}}}]{Saito11}
\bibinfo{author}{\bibfnamefont{S.}~\bibnamefont{{Saito}}},
  \bibinfo{author}{\bibfnamefont{M.}~\bibnamefont{{Takada}}}, \bibnamefont{and}
  \bibinfo{author}{\bibfnamefont{A.}~\bibnamefont{{Taruya}}},
  \bibinfo{journal}{\prd} \textbf{\bibinfo{volume}{83}},
  \bibinfo{pages}{043529} (\bibinfo{year}{2011}), \eprint{1006.4845}.

\bibitem[{\citenamefont{{Jeong} and {Komatsu}}(2006)}]{Jeong06}
\bibinfo{author}{\bibfnamefont{D.}~\bibnamefont{{Jeong}}} \bibnamefont{and}
  \bibinfo{author}{\bibfnamefont{E.}~\bibnamefont{{Komatsu}}},
  \bibinfo{journal}{\apj} \textbf{\bibinfo{volume}{651}}, \bibinfo{pages}{619}
  (\bibinfo{year}{2006}), \eprint{arXiv:astro-ph/0604075}.

\bibitem[{\citenamefont{{Nishimichi} et~al.}(2009)\citenamefont{{Nishimichi},
  {Shirata}, {Taruya}, {Yahata}, {Saito}, {Suto}, {Takahashi}, {Yoshida},
  {Matsubara}, {Sugiyama} et~al.}}]{Nishimichi09}
\bibinfo{author}{\bibfnamefont{T.}~\bibnamefont{{Nishimichi}}},
  \bibinfo{author}{\bibfnamefont{A.}~\bibnamefont{{Shirata}}},
  \bibinfo{author}{\bibfnamefont{A.}~\bibnamefont{{Taruya}}},
  \bibinfo{author}{\bibfnamefont{K.}~\bibnamefont{{Yahata}}},
  \bibinfo{author}{\bibfnamefont{S.}~\bibnamefont{{Saito}}},
  \bibinfo{author}{\bibfnamefont{Y.}~\bibnamefont{{Suto}}},
  \bibinfo{author}{\bibfnamefont{R.}~\bibnamefont{{Takahashi}}},
  \bibinfo{author}{\bibfnamefont{N.}~\bibnamefont{{Yoshida}}},
  \bibinfo{author}{\bibfnamefont{T.}~\bibnamefont{{Matsubara}}},
  \bibinfo{author}{\bibfnamefont{N.}~\bibnamefont{{Sugiyama}}},
  \bibnamefont{et~al.}, \bibinfo{journal}{Publ. Astron. Soc. Japan}
  \textbf{\bibinfo{volume}{61}}, \bibinfo{pages}{321} (\bibinfo{year}{2009}),
  \eprint{0810.0813}.

\bibitem[{\citenamefont{{Jenkins} et~al.}(1998)\citenamefont{{Jenkins},
  {Frenk}, {Pearce}, {Thomas}, {Colberg}, {White}, {Couchman}, {Peacock},
  {Efstathiou}, and {Nelson}}}]{Jenkins98}
\bibinfo{author}{\bibfnamefont{A.}~\bibnamefont{{Jenkins}}},
  \bibinfo{author}{\bibfnamefont{C.~S.} \bibnamefont{{Frenk}}},
  \bibinfo{author}{\bibfnamefont{F.~R.} \bibnamefont{{Pearce}}},
  \bibinfo{author}{\bibfnamefont{P.~A.} \bibnamefont{{Thomas}}},
  \bibinfo{author}{\bibfnamefont{J.~M.} \bibnamefont{{Colberg}}},
  \bibinfo{author}{\bibfnamefont{S.~D.~M.} \bibnamefont{{White}}},
  \bibinfo{author}{\bibfnamefont{H.~M.~P.} \bibnamefont{{Couchman}}},
  \bibinfo{author}{\bibfnamefont{J.~A.} \bibnamefont{{Peacock}}},
  \bibinfo{author}{\bibfnamefont{G.}~\bibnamefont{{Efstathiou}}},
  \bibnamefont{and} \bibinfo{author}{\bibfnamefont{A.~H.}
  \bibnamefont{{Nelson}}}, \bibinfo{journal}{\apj}
  \textbf{\bibinfo{volume}{499}}, \bibinfo{pages}{20} (\bibinfo{year}{1998}),
  \eprint{arXiv:astro-ph/9709010}.

\bibitem[{\citenamefont{{Hamana} et~al.}(2003)\citenamefont{{Hamana}, {Kayo},
  {Yoshida}, {Suto}, and {Jing}}}]{Hamana03}
\bibinfo{author}{\bibfnamefont{T.}~\bibnamefont{{Hamana}}},
  \bibinfo{author}{\bibfnamefont{I.}~\bibnamefont{{Kayo}}},
  \bibinfo{author}{\bibfnamefont{N.}~\bibnamefont{{Yoshida}}},
  \bibinfo{author}{\bibfnamefont{Y.}~\bibnamefont{{Suto}}}, \bibnamefont{and}
  \bibinfo{author}{\bibfnamefont{Y.~P.} \bibnamefont{{Jing}}},
  \bibinfo{journal}{Mon. Not. Roy. Astron. Soc.}
  \textbf{\bibinfo{volume}{343}}, \bibinfo{pages}{1312} (\bibinfo{year}{2003}),
  \eprint{arXiv:astro-ph/0305187}.

\bibitem[{\citenamefont{{Linder}}(2005)}]{Linder05}
\bibinfo{author}{\bibfnamefont{E.~V.} \bibnamefont{{Linder}}},
  \bibinfo{journal}{\prd} \textbf{\bibinfo{volume}{72}},
  \bibinfo{pages}{043529} (\bibinfo{year}{2005}),
  \eprint{arXiv:astro-ph/0507263}.

\bibitem[{\citenamefont{{Blake}
  et~al.}(2011{\natexlab{b}})\citenamefont{{Blake}, {Brough}, {Colless},
  {Contreras}, {Couch}, {Croom}, {Davis}, {Drinkwater}, {Forster}, {Gilbank}
  et~al.}}]{Blake11a}
\bibinfo{author}{\bibfnamefont{C.}~\bibnamefont{{Blake}}},
  \bibinfo{author}{\bibfnamefont{S.}~\bibnamefont{{Brough}}},
  \bibinfo{author}{\bibfnamefont{M.}~\bibnamefont{{Colless}}},
  \bibinfo{author}{\bibfnamefont{C.}~\bibnamefont{{Contreras}}},
  \bibinfo{author}{\bibfnamefont{W.}~\bibnamefont{{Couch}}},
  \bibinfo{author}{\bibfnamefont{S.}~\bibnamefont{{Croom}}},
  \bibinfo{author}{\bibfnamefont{T.}~\bibnamefont{{Davis}}},
  \bibinfo{author}{\bibfnamefont{M.~J.} \bibnamefont{{Drinkwater}}},
  \bibinfo{author}{\bibfnamefont{K.}~\bibnamefont{{Forster}}},
  \bibinfo{author}{\bibfnamefont{D.}~\bibnamefont{{Gilbank}}},
  \bibnamefont{et~al.}, \bibinfo{journal}{ArXiv e-prints}
  (\bibinfo{year}{2011}{\natexlab{b}}), \eprint{1104.2948}.

\bibitem[{\citenamefont{{Jennings} et~al.}(2011)\citenamefont{{Jennings},
  {Baugh}, and {Pascoli}}}]{Jennings11}
\bibinfo{author}{\bibfnamefont{E.}~\bibnamefont{{Jennings}}},
  \bibinfo{author}{\bibfnamefont{C.~M.} \bibnamefont{{Baugh}}},
  \bibnamefont{and}
  \bibinfo{author}{\bibfnamefont{S.}~\bibnamefont{{Pascoli}}},
  \bibinfo{journal}{Mon. Not. Roy. Astron. Soc.}
  \textbf{\bibinfo{volume}{410}}, \bibinfo{pages}{2081} (\bibinfo{year}{2011}),
  \eprint{1003.4282}.

\end{thebibliography}

\end{document}